\documentclass[reqno,12pt]{article}
\usepackage{amsfonts,latexsym,amscd,amssymb}
\usepackage{graphicx}
\usepackage{amsmath}
\usepackage{epsfig}

\newcommand{\stackconj}[1]{\;\stackrel{?[{#1}]}{=\joinrel=}\;}
\newcommand{\stacktick}[1]{\;\stackrel{\checkmark[{#1}]}{=\joinrel=}\;}

\newbox\shell
\newcommand{\dia}[2]{\setbox\shell=\hbox{\begin{picture}(180,220)(-90,-110)#1
\put(-90,-90){\makebox(180,220)[b]{\large #2}}\end{picture}}\dimen0=\ht
\shell\multiply\dimen0by7\divide\dimen0by16\raise-\dimen0\box\shell\hfill}
\newcommand{\vtx}{\circle*{10}}

\begin{document}

\title{\bf Elliptic integral evaluations of Bessel moments}

\author{David H.\ Bailey\thanks{Lawrence Berkeley National Laboratory,
Berkeley, CA 94720, USA,
{\tt dhbailey@lbl.gov}.
Supported in part by the Director, Office of Computational and Technology
Research, Division of Mathematical, Information and Computational
Sciences, U.S.\ Department of Energy, under contract number
DE-AC02-05CH11231.},
Jonathan M.\ Borwein\thanks{Faculty of Computer Science, Dalhousie University,
Halifax, NS, B3H 2W5, Canada,
{\tt jborwein@cs.dal.ca}.
Supported in part by NSERC and the Canada Research Chair Programme.},
David Broadhurst\thanks{Department of Physics and Astronomy,
The Open University, Milton Keynes, MK7 6AA, United Kingdom,
{\tt D.Broadhurst@open.ac.uk}.}, and
M.L.\ Glasser\thanks{Department of Physics,
Clarkson University, Potsdam, NY 13699-5820, USA,
{\tt laryg@clarkson.edu}.}}

\date{\today}

\maketitle

\abstract{We record what is known about the closed forms for
various Bessel function moments arising in quantum field theory,
condensed matter theory and other parts of mathematical physics.
More generally, we develop formulae for integrals of products of
six or fewer Bessel functions. In consequence,
we are able to discover and prove closed forms for
$c_{n,k}:=\int_0^\infty t^k K_0^n(t)\,{\rm d}t$
with integers $n=1,2,3,4$ and $k\ge0$, obtaining
new results for the even moments $c_{3,2k}$ and $c_{4,2k}$.
We also derive new closed forms for the odd moments
$s_{n,2k+1}:=\int_0^\infty t^{2k+1}I_0^{}(t)\,K_0^{n-1}(t)\,{\rm d}t$
with $n=3,4$ and for
$t_{n,2k+1}:=\int_0^\infty t^{2k+1}I_0^2(t)\,K_0^{n-2}(t)\,{\rm d}t$
with $n=5$, relating the latter to Green functions on hexagonal,
diamond and cubic lattices.
We conjecture the values of $s_{5,2k+1}$,
make substantial progress on the evaluation
of $c_{5,2k+1}$, $s_{6,2k+1}$ and $t_{6,2k+1}$
and report more limited progress regarding
$c_{5,2k}$, $c_{6,2k+1}$ and $c_{6,2k}$.

In the process, we obtain 8 conjectural evaluations, each of
which has been checked to 1200 decimal places. One of
these lies deep in 4-dimensional quantum field theory and two are
probably provable by delicate combinatorics. There remains a hard
core of five conjectures whose proofs would be most instructive, to
mathematicians and physicists alike.}

\newpage

\section{Introduction}

We give representations of the \emph{vacuum}~\cite{davtau,Lap21} and
\emph{sunrise}~\cite{anydim,groote,sunrise} diagrams
\begin{eqnarray}
V_n(a_1,\ldots,a_n)&:=&
\int_0^\infty t\left(\prod_{j=1}^n K_0(a_j t)\right)\,{\rm d}t
\label{vacuum}\\
S_{n+1}(a_1,\ldots,a_n,w)&:=&
\int_0^\infty t\left(\prod_{j=1}^{n}K_0(a_j t)\right)J_0(w t)\,{\rm d}t
\label{sunrise}
\end{eqnarray}
where all the arguments of $V_n$ and all
but the last argument of $S_{n+1}$ are real and positive.
Here and below $I_0, J_0$ and $K_0$ are the
conventional \emph{Bessel functions} of order zero
as in~\cite[Chapter 15]{AandS}.
These integrals occur in quantum field theories in two spacetime
dimensions, where we do not need to regularize ultraviolet divergences.
Numbers generated by them are expected to occur in
the finite parts of integrals from Feynman diagrams in four spacetime
dimensions. To be concrete, we illustrate $V_3$ and $S_4$ as follows:

\mbox{\hspace{1cm}}\hfill
\dia{
\put(-50,0){\line(1,0){100}}
\put(0,0){\circle{100}}
\put(50,0){\vtx}
\put(-50,0){\vtx}
}{$V_3$}
\dia{
\put(-100,0){\line(1,0){200}}
\put(0,0){\circle{100}}
\put(50,0){\vtx}
\put(-50,0){\vtx}
}{$S_4$}
\mbox{\hspace{1cm}}\par

By casting Bessel's differential equation in the form
\[\left(\frac{1}{a}+\frac{{\rm d}}{{\rm d}a}\right)
\frac{{\rm d}}{{\rm d}a}\,K_0(a t)=t^2K_0(a t)\]
and applying the corresponding differential operator
to~(\ref{vacuum}), we may increase the exponent of $t$ in
the integrand by steps of 2.
But to obtain even powers of $t$, we need to start with
\begin{equation}
\overline{V}_n(a_1,\ldots,a_n):=
\int_0^\infty \left(\prod_{j=1}^n K_0(a_j t)\right)\,{\rm d}t
\label{novacuum}
\end{equation}
which plays no obvious role in quantum field theory.
To evaluate the latter form, we found it useful
to regard the Fourier transform
\begin{equation}
\overline{S}_{n+1}(a_1,\ldots,a_n,w):=
\frac{1}{\pi}
\int_0^\infty \left(\prod_{j=1}^{n}K_0(a_j t)\right)\cos(w t)\,{\rm d}t\
\label{nosunrise}
\end{equation}
as an analogue of~(\ref{sunrise}).

We shall be especially interested in the moments
\begin{equation}
c_{n,k}:=\int_0^\infty t^k K_0^n(t)\,{\rm d}t
\label{moments}
\end{equation}
for integers $n\ge1$ and $k\ge0$,
as studied in~\cite{bbc06,bbbc07} and~\cite{bsalvy,ouvry}.

In~\cite{bbbc07} these moments arose in the study of Ising-type integrals
\[C_{n,k}=\frac{1}{n!}\int_0^{\infty}\cdots\int_0^{\infty}
\frac{{\rm d}x_1 \ {\rm d}x_2 \cdots \ {\rm d}x_n}
{(\cosh x_1 + \dots + \cosh x_n)^{k+1}}\]
which are linked by
\begin{equation}
C_{n,k}=\frac{2^n}{n!\,k!}\,c_{n,k}\,.
\label{cnkbess}
\end{equation}

In~\cite{bsalvy,ouvry} it is proven that for fixed $n$ these
moments satisfy a linear recursion for which a simple algorithm exists
with coefficients polynomial in $k$.
For example, for $n=1$ and $2$ one easily obtains the closed forms
\begin{equation}
c_{1,k}=2^{k-1}\,\Gamma^2\left(\frac{k+1}{2}\right)\,\mbox{ and }\,
c_{2,k}=\frac{\sqrt{\pi}\;\Gamma^3\left(\frac{k+1}{2}\right)}
{4\,\Gamma\left(\frac{k}{2}+1\right)}
\label{closed}
\end{equation}
and for $n=3$ and $4$ we obtain the recursions
\begin{eqnarray}
(k+1)^4c_{3,k}-2(5k^2+20k+21)c_{3,k+2}+9c_{3,k+4}&=&0
\label{salvy3}\\
(k+1)^5c_{4,k}-4(k+2)(5k^2+20k+23)c_{4,k+2}+64(k+3)c_{4,k+4}&=&0\,.
\label{salvy4}
\end{eqnarray}
These recursion formulae may be written quite compactly as
\begin{equation}
\sum_{i=0}^M\, (-1)^i\,p_{n,i}(k+i+1)\,c_{n,k+2i}= 0
\label{littlecrec}
\end{equation}
where $M = \lfloor (n+1)/2 \rfloor$.
For instance, for $n=5$ and $6$, we have
\begin{equation}
\begin{array}{ll}
p_{5,0}(x) = x^6 & p_{6,0}(x) = x^7 \\
p_{5,1}(x) = 35 x^4 + 42 x^2 + 3 & p_{6,1}(x) =
x(56x^4+112x^2+24) \\
p_{5,2}(x) = 259 x^2 + 104 & p_{6,2}(x) = x(784x^2+944) \\
p_{5,3}(x) = 225 & p_{6,3}(x) = 2304x. \\
\end{array}
\label{crec56}
\end{equation}

The same recursions apply to the moments
\begin{equation}
s_{n,k}:=\int_0^\infty t^k I_0(t)K_0^{n-1}(t)\,{\rm d}t
\label{smoments}
\end{equation}
for integers $n\ge3$ and $k\ge0$ and to
\begin{equation}
t_{n,k}:=\int_0^\infty t^k I_0^2(t)K_0^{n-2}(t)\,{\rm d}t
\label{tmoments}
\end{equation}
for integers $n\ge5$ and $k\ge0$.

\section{Two Bessel functions}

The transforms
\begin{eqnarray}
S_2(a,w)&:=&
\int_0^\infty t\,K_0(a t)J_0(w t)\,{\rm d}t
=\frac{1}{a^2+w^2}
\label{ess2}\\
\overline{S}_2(a,w)&:=&
\frac{1}{\pi}\int_0^\infty K_0(a t)\cos(w t)\,{\rm d}t
=\frac{1}{2\sqrt{a^2+w^2}}
\label{noess2}
\end{eqnarray}
give $V_1(a)=1/a^2$ and
$\overline{V}_1(a)/\pi=1/(2a)$ at $w=0$.
The distributions
\begin{eqnarray}
\int_0^\infty v J_0(v t_1)J_0(v t_2)\,{\rm d}v&=&2\delta(t_1^2-t_2^2)
\label{qftdelta}\\
\frac{2}{\pi}\int_0^\infty \cos(v t_1)\cos(v t_2)\,{\rm d}v&=&
\delta(t_1+t_2)+\delta(t_1-t_2)
\label{cosdelta}
\end{eqnarray}
then lead to the evaluations
\begin{eqnarray}
V_2(a,b)&:=&\int_0^\infty t\,K_0(a t)K_0(b t)\,{\rm d}t
=\int_0^\infty w S_2(a,w)S_2(b,w)\,{\rm d}w
\nonumber\\
&=&\frac{\log(a/b)}{a^2-b^2}
\label{vee2}\\
\overline{V}_2(a,b)&:=&\int_0^\infty K_0(a t)K_0(b t)\,{\rm d}t
=2\pi\int_0^\infty \overline{S}_2(a,w)\overline{S}_2(b,w)\,{\rm d}w
\nonumber\\
&=&\frac{\pi}{a+b}\,{\bf K}\left(\frac{a-b}{a+b}\right)
\label{novee2}
\end{eqnarray}
with a \emph{complete elliptic integral} of the first kind,
\begin{equation}
{\bf K}(k):=\int_0^{\pi/2}\frac{1}{\sqrt{1-k^2\sin^2\phi}}\,{\rm d}\phi\,,
\label{Kk}
\end{equation}
appearing in~(\ref{novee2}) and a limit intended in~(\ref{vee2}) when $a=b$.
We shall need to refer to the \emph{complementary}
integral ${\bf K}^\prime(k):={\bf K}(k^\prime)$, with $k^\prime:=\sqrt{1-k^2}$.
In the case $a\ge b$, this provides a compact alternative form
of~(\ref{novee2}),
\begin{equation}
\overline{V}_2(a,b)=\frac{\pi}{2a}\,{\bf K}^\prime(b/a)\,,
\label{novee2alt}
\end{equation}
obtained by the Landen~\cite{agmhist} transformation in~\cite[17.3.29]{AandS}.

\section{Three Bessel functions}

We follow K\"all\'en~\cite{kallen,KS}
by constructing $S_3(a,b,w)$ from its
discontinuity across the cut in the $w^2$ plane with branch point
at $w^2=-(a+b)^2$, obtaining
\[S_3(a,b,w)=\int_0^\infty t\,K_0(a t)K_0(b t)J_0(w t)\,{\rm d}t
=\int_{a+b}^\infty\frac{2v\,D_3(a,b,v)}{v^2+w^2}\,{\rm d}v\]
with a discontinuity
\[D_3(a,b,c)=\frac{1}{\sqrt{(a+b+c)(a-b+c)(a+b-c)(a-b-c)}}\]
that is completely symmetric in its 3 arguments.
The $v$-integral is easily performed, to give
\begin{equation}
S_3(a,b,w)=2\,{\rm arctanh}
\left(\sqrt{\frac{w^2+(a-b)^2}{w^2+(a+b)^2}}\right)
D_3(a,b,{\rm i}w)
\label{ess3}
\end{equation}
and in particular the \emph{on-shell} value~\cite{groote}
\[s_{3,1}=S_3(1,1,{\rm i}):=\int_0^\infty t\,I_0(t)K_0^2(t)\,{\rm d}t
=L_{-3}(1)=\frac{\pi}{3\sqrt{3}}\]
where
\[L_{-3}(s):=
\sum_{n=1}^\infty\frac{\chi_{-3}(n)}{n^s}
=\sum_{k=0}^\infty\left(\frac{1}{(3k+1)^s}
-\frac{1}{(3k+2)^s}\right)\]
is the Dirichlet $L$-function with the real character $\chi_{-3}(n)$ given
by the Legendre--Jacobi--Kronecker symbol $(D|n)$ for discriminant $D=-3$.

\paragraph{Nomenclature.} We shall refer to the construction
of a Feynman amplitude from its discontinuity across a cut as a
\emph{dispersive} calculation. Kramers~\cite{kramers} and Kronig~\cite{kronig}
founded this approach in studies of the dispersion of light, in the 1920's.
In the 1950's, the utility of dispersive methods was recognized in
particle physics~\cite{jostlutt,KS,kampen}. Cutkosky~\cite{cutkosky}
turned them into a calculus that became a routine part of
the machinery of quantum field theory. Barton~\cite{Barton} has given
a scholarly and instructive introduction to these techniques.

\subsection{The odd moments $s_{3,2k+1}$}

By differentiating
\[S_3(1,1,{\rm i}x)=\frac{2\,{\rm arcsin}(x/2)}{x\sqrt{4-x^2}}\]
before setting $x=1$, we evaluate
\[s_{3,3}:=\int_0^\infty t^3 I_0(t)K_0^2(t)\,{\rm d}t=
\left.\left(\frac{1}{x}+\frac{{\rm d}}{{\rm d}x}\right)
\frac{{\rm d}S_3(1,1,{\rm i}x)}{{\rm d}x}\right|_{x=1}
=\frac43\,s_{3,1}\]
and are then able to solve the recursion relation~\cite{bsalvy,ouvry}
for $s_{3,2k+1}$ by the \emph{closed form}
\begin{equation}
s_{3,2k+1}=\frac{\pi}{3\sqrt3}\left(\frac{2^k{}k!}{3^k}\right)^2a_k
\label{ess3k}
\end{equation}
with integers
\begin{equation}
a_k=\sum_{j=0}^k{k\choose j}^2{2j\choose j}\,.
\label{A2893}
\end{equation}

Integer sequence~(\ref{A2893}) begins
\begin{equation}
1,\,3,\,15,\,93,\,639,\,4653,\,35169,\,272835,\,2157759,\,17319837,\,
140668065,\,1153462995
\label{A2893s}
\end{equation}
and is recorded\footnote{See {\tt
http://www.research.att.com/\~{ }njas/sequences/A002893}~.}
as entry A2893 of the on-line version of~\cite{EIS},
which gives the recursion
\begin{equation}
(k+1)^2a_{k+1}-(10k^2+10k+3)a_k+9k^2a_{k-1}=0
\label{A2893r}
\end{equation}
and the generating function
\begin{equation}
I_0^3(2t)=\sum_{k=0}^\infty a_k\left(\frac{t^k}{k!}\right)^2\,.
\label{A2893g}
\end{equation}
We have verified that recursion~(\ref{A2893r}) reproduces the recursion
for~(\ref{ess3k}), which has the same form as for the odd
moments in~(\ref{salvy3}). We note that integers $a_k$ were encountered
in studies of cooperative phenomena in crystals~\cite{crystals}
and also in studies of matrices~\cite{matrices} with entries 0 or 1.
In~\cite[Prop.\ 2]{hexagonal}, they are related to enumeration of closed
walks in a two-dimensional hexagonal lattice. They also appear in an
enumeration of Feynman diagrams~\cite[Table 2]{partons} in
quantum chromodynamics, via the constrained sum~\cite{SIAM17}
\[a_k=\sum_{p+q+r=k}\left(\frac{k!}{p!\,q!\,r!}\right)^2\]
that results from the Taylor expansion of $I_0^3$.

It is notable that $I_0^3$ provides a generating function
for moments of $I_0^{}K_0^2$. In Section~4 we shall show that
$I_0^4$ generates moments of $I_0^{}K_0^3$.

\subsection{The odd moments $c_{3,2k+1}$}

Next, we construct
\begin{eqnarray}
V_3(a,b,c)&:=&\int_0^\infty t\,K_0(a t)K_0(b t)K_0(c t)\,{\rm d}t
=\int_0^\infty w S_3(a,b,w)S_2(c,w)\,{\rm d}w
\nonumber\\
&=&\frac{L_3(a,b,c)+L_3(b,c,a)+L_3(c,a,b)}{4}D_3(a,b,c)
\label{vee3}
\end{eqnarray}
with a dilogarithmic function
\[L_3(a,b,c)
:={\rm Li}_2\left(
\frac{(a^2+b^2-c^2)D_3(a,b,c)+1}{(a^2+b^2-c^2)D_3(a,b,c)-1}\right)
-{\rm Li}_2\left(
\frac{(a^2+b^2-c^2)D_3(a,b,c)-1}{(a^2+b^2-c^2)D_3(a,b,c)+1}\right)\]
computed by Davydychev and Tausk~\cite{davtau}.
Setting $a=b=c=1$, we obtain
\[c_{3,1}=V_3(1,1,1)
=\frac34\,L_{-3}(2)
=\frac{1}{9}\sum_{n=0}^\infty\left(\frac{-1}{27}\right)^n
\sum_{k=1}^5\frac{v_k}{(6n+k)^2}\]
where the vector of coefficients $v=[9,-9,-12,-3,1]$
was discovered (and proven) in the
course of investigation of 3-loop vacuum diagrams in 4
dimensions~\cite{3loop}.

Similarly,
\[c_{3,3}=L_{-3}(2)-\frac{2}{3}\]
may be obtained by suitable differentiations of~(\ref{vee3}).
Then higher moments $c_{3,2k+1}$ with $k>1$ may be obtained
by using~(\ref{salvy3}). Because of the mixing of $L_{-3}(2)$
with unity, we were unable to write a closed form for their
rational coefficients in $c_{3,2k+1}$.

\subsection{The even moments $c_{3,2k}$}

For even moments, we lack the dispersion relations~\cite{Barton}
of quantum field theory and so fall back on the general \emph{Aufbau}
\begin{equation}
\overline{S}_{m+n+1}(a_1,\ldots a_m,b_1,\ldots b_n,w)
=\int_{-\infty}^\infty
\overline{S}_{m+1}(a_1,\ldots a_m,v)
\overline{S}_{n+1}(b_1,\ldots b_n,v+w)\,{\rm d}v
\label{Aufbau}
\end{equation}
which follows from the distribution
\[\frac{1}{\pi}\int_{-\infty}^\infty \cos(v t_1)\cos((v+w)t_2)\,{\rm d}v
=\delta(t_1+t_2)\cos(w t_2)+\delta(t_1-t_2)\cos(w t_2)\]
obtained from~(\ref{cosdelta}) and the expansion
$\cos((v+w)t_2)=\cos(v t_2)\cos(w t_2)-\sin(v t_2)\sin(w t_2)$.

In particular, by setting $m=n=1$ in~(\ref{Aufbau}), we evaluate
\[\overline{S}_3(a,b,w):=
\frac{1}{\pi}\int_0^\infty K_0(a t)K_0(b t)\cos(w t)\,{\rm d}t
=\int_{-\infty}^\infty
\overline{S}_2(a,v)
\overline{S}_2(b,v+w)\,{\rm d}v\]
as the complete elliptic integral
\begin{equation}
\overline{S}_3(a,b,w)=
\frac14\int_{-\infty}^\infty\frac{{\rm d}v}
{\sqrt{(a^2+v^2)(b^2+(v+w)^2)}}=
\frac{{\bf K}\left(\sqrt{\frac{w^2+(a-b)^2}{w^2+(a+b)^2}}\right)}
{\sqrt{w^2+(a+b)^2}}
\label{noess3}
\end{equation}
and recover the previous
result~(\ref{novee2}) for $\overline{V}_2(a,b)=\pi\overline{S}_3(a,b,0)$
by setting $w=0$.

Next, by setting $m=2$ and $n=1$ in~(\ref{Aufbau}), we write
\[\overline{S}_4(a,b,c,w):=
\frac{1}{\pi}\int_0^\infty K_0(a t)K_0(b t)K_0(c t)\cos(w t)\,{\rm d}t
=\int_{-\infty}^\infty
\overline{S}_3(a,b,v)
\overline{S}_2(c,v+w)\,{\rm d}v\]
as an integral over an elliptic integral:
\begin{equation}
\overline{S}_4(a,b,c,w)=
\frac12\int_{-\infty}^\infty
\frac{{\bf K}\left(\sqrt{\frac{v^2+(a-b)^2}{v^2+(a+b)^2}}\right)}
{\sqrt{(v^2+(a+b)^2)((v+w)^2+c^2)}}\,{\rm d}v
\label{noess4}
\end{equation}
and at $w=0$ obtain
\begin{equation}
\overline{V}_3(a,b,c)
:=\int_0^\infty K_0(a t)K_0(b t)K_0(c t)\,{\rm d}t
=\pi\int_0^\infty
\frac{{\bf K}\left(\sqrt{\frac{v^2+(a-b)^2}{v^2+(a+b)^2}}\right)}
{\sqrt{(v^2+(a+b)^2)(v^2+c^2)}}\,{\rm d}v\,.
\label{novee3}
\end{equation}

Remarkably, the integral~(\ref{novee3}) may be evaluated
by exploiting a more general identity given in W.N.\ Bailey's
second paper on infinite integrals involving Bessel
functions~\cite{bailey1936}. Without loss of generality,
we assume that $c\ge b\ge a>0$ and define
\[k_\pm:=\frac{\sqrt{(c+a)^2-b^2}\pm\sqrt{(c-a)^2-b^2}}{2c}\,,
\quad k_\pm^\prime:=\sqrt{1-k_\pm^2}\,.\]
Then our result may written as
\begin{equation}
\frac{2c}{\pi}\,\overline{V}_3(a,b,c)=
{\bf K}(k_-)\,{\bf K}(k_+^\prime)+
{\bf K}(k_+)\,{\bf K}(k_-^\prime)\,.
\label{djbwnb}
\end{equation}
We remark that when $c>a+b$ each term in~(\ref{djbwnb}) is real;
otherwise, each is the complex conjugate of the other. The
form of $k_\pm$ comes from Bailey's conditions
$k_+k_-=a/c$ and $k^\prime_+k^\prime_-=b/c$.

We proved~(\ref{djbwnb}) by
setting $\nu=\rho=0$ in Bailey's equation~(3.3), to obtain
\[\int_0^\infty I_\mu(a t)K_0(b t)K_0(c t)\,{\rm d}t=
\frac{1}{4c}W_\mu(k_+)W_\mu(k_-)\]
with a hypergeometric series
\[W_\mu(k):=\sum_{n=0}^\infty
\frac{\Gamma^2\left(n+\frac{1+\mu}{2}\right)k^{2n+\mu}}
{\Gamma(n+1+\mu)n!}=
\frac{\sqrt{\pi}\,\Gamma\left(\frac{1+\mu}{2}\right)}{(1-k^2)^{1/4}}
P^{-\mu/2}_{-1/2}\left(\frac{2-k^2}{2\sqrt{1-k^2}}\right)\]
where $P$ is the Legendre function defined in~\cite[8.1.2]{AandS}.
Then~(\ref{djbwnb}) follows from the expansions
\begin{eqnarray*}
I_\mu(x)&=&I_0(x)-\mu K_0(x)+{\rm O}(\mu^2)\\
W_\mu(k)&=&2{\bf K}(k)-\mu\pi{\bf K}\left(\sqrt{1-k^2}\right)+{\rm O}(\mu^2)
\end{eqnarray*}
where the derivative of $W_\mu(k)$ at $\mu=0$ is obtained
by setting $a=b=\frac12$ and $z=1-k^2$ in~\cite[15.3.10]{AandS}.

Specializing~(\ref{djbwnb}) to the case $a=b=1$ we obtain
\begin{equation}
\frac{2c}{\pi}\int_0^\infty K_0^2(t)K_0(c t)\,{\rm d}t
=A(2/c)=B(c/2)
\label{AandB}
\end{equation}
with the choice of a sum of squares or a product in the functions
\begin{eqnarray}
A(x)&:=&
{\bf K}^2\left(\frac{\sqrt{1+x}-\sqrt{1-x}}{2}\right)+
{\bf K}^2\left(\frac{\sqrt{1+x}+\sqrt{1-x}}{2}\right)
\label{Aform}\\
B(x)&:=&2x
{\bf K}\left(\frac{\sqrt{1+x}-\sqrt{1-x}}{2}\right)
{\bf K}\left(\frac{\sqrt{1+x}+\sqrt{1-x}}{2}\right)
\label{Bform}
\end{eqnarray}
with~(\ref{Aform}) coming directly from~(\ref{djbwnb}).
The identity $A(x)=B(1/x)$ may be proven by showing that
$A(1-y)$ and $B(1/(1-y))$
satisfy the same third-order
differential equation and have Taylor series about $y=0$
that agree in their first 3 terms.

Alternative one may
use transformations of the \emph{Meijer G-functions}
\begin{eqnarray}
2\sqrt{\pi}A(x)&=&G^{23}_{33}\left(\frac{1}{x^2}\left|
\begin{array}{ccc}1&1&1\\
\frac12&\frac12&\frac12\end{array}\right.\right)
=G^{32}_{33}\left(x^2\left|
\begin{array}{ccc}\frac12&\frac12&\frac12\\
0&0&0\end{array}\right.\right)
\label{Ameijer}\\
2\sqrt{\pi}B(x)&=&G^{23}_{33}\left(x^2\left|
\begin{array}{ccc}1&1&1\\
\frac12&\frac12&\frac12\end{array}\right.\right)
=x\,G^{23}_{33}\left(x^2\left|
\begin{array}{ccc}\frac12&\frac12&\frac12\\
0&0&0\end{array}\right.\right)
\label{Bmeijer}
\end{eqnarray}
in the notation of ~\cite[Vol.\ 3]{prud}. This
provides an inversion formula, 8.2.1.14,
used in~(\ref{Ameijer}) and a multiplication formula, 8.2.1.15,
used in~(\ref{Bmeijer}).

\paragraph{Remark.} Our forms,~(\ref{Aform}) and~(\ref{Bform}),
for the Bessel integral~(\ref{AandB}), were
tabulated, without proof, in identities 2.16.46.4 and 2.16.46.5
of~\cite[Vol.\ 2]{prud}.
Our proof of the more general identity~(\ref{djbwnb})
came from following a reference to Bailey's work,
given in Section 7.14.2.43
of the Bateman project~\cite[Vol.\ 2]{erdelyi}.

\medskip

Setting $c=1$ in~(\ref{AandB}), we obtain
\begin{equation}
c_{3,0}=\frac{\pi}{2}\,K_3K_3^\prime=
\frac{3\,\Gamma^6\left(\frac13\right)}{32\pi 2^{2/3}}
\label{c30}
\end{equation}
with the product of $K_3={\bf K}(k_3)$
and $K_3^\prime=\sqrt{3}K_3$
obtained from~(\ref{Bform}), at $x=\frac12$, where the third
\emph{singular value}~\cite{agm}
\[k_3=\frac{\sqrt3-1}{2\sqrt2}=\sin(\pi/12)\]
results. Moreover, Bessel's differential equation yields
\begin{equation}
c_{3,2}=\frac{\pi}{2}
\left.\left(\frac{1}{c}+\frac{{\rm d}}{{\rm d}c}\right)
\frac{{\rm d}}{{\rm d}c}
\frac{B(c/2)}{c}\right|_{c=1}=
\frac{\Gamma^6\left(\frac{1}{3}\right)}{96\pi 2^{2/3}}
-\frac{4\pi^5 2^{2/3}}{9\,\Gamma^6\left(\frac{1}{3}\right)}
\label{c32}
\end{equation}
upon use of the evaluations
\begin{eqnarray*}
{\bf E}(\sin(\pi/12))&=&\frac{K_3^\prime+3K_3}{6}+\frac{\pi}{4K_3^\prime}\\
{\bf E}(\cos(\pi/12))&=&\frac{K_3^\prime-K_3}{2}+\frac{\pi}{4K_3}
\end{eqnarray*}
of complete elliptic integrals of the second kind,
recorded in~\cite[p.\ 28]{agm}
and first found by Legendre. Prior to finding this proof,
we discovered~(\ref{c32}) in the more palatable form
\begin{equation}
c_{3,2}=
\frac19\,c_{3,0}-\frac{\pi^4}{24}\,c_{3,0}^{-1}
\label{c32pslq}
\end{equation}
by using PSLQ~\cite{pslq,2level,expm1} in a manner suggested
by previous discoveries in quantum field theory, as described in Section~5.

\subsection{Continued fraction}

We recall that
\begin{equation}
9\,\frac{c_{3,2}}{c_{3,0}}=
\cfrac{9\cdot1^4}{d(1)-\cfrac{9\cdot3^4}{\ddots-\cfrac{9\cdot(2N-1)^4}
{d(N)-\dots}}}
\label{contfrac3e}
\end{equation}
where $d(N):=40N^2+2$ was derived in~\cite{bbbc07}. Hence,
dividing~(\ref{c32pslq}) by $c_{3,0}$, we obtain
a neat continued fraction for
\begin{equation}
1-2^{7/3}9\left(\frac
{\Gamma\left(\frac23\right)}
{\Gamma\left(\frac13\right)}\right)^6=9\,\frac{c_{3,2}}{c_{3,0}}\,.
\label{contfrac}
\end{equation}

\subsection{Double integrals}

Inspection of~\cite{bbbc07} also reveals that we now
have evaluated the integrals
\begin{eqnarray}
\int_0^\infty\int_0^\infty\frac{{\rm d}x\,{\rm d}y}
{\sqrt{(1+x^2)(1+y^2)(1+(x+y)^2)}}
&=&\frac{2}{3}\,K_3K_3^\prime
\label{DJB3b}\\
\int_0^\infty\int_0^\infty\frac{{\rm d}x\,{\rm d}y}
{\sqrt{(1+x^2)(1+y^2)(1+(x-y)^2)}}
&=&\frac{4}{3}\,K_3K_3^\prime
\label{DJB3c}
\end{eqnarray}
The first integral occurs in a formula for $4c_{3,0}/(3\pi)$
in~\cite{bbbc07}.
To evaluate the second integral, we note that it is twice the value
obtained from the simplex $x>y>0$. The transformation
$x=y+z$ then proves that~(\ref{DJB3c}) is twice~(\ref{DJB3b}).

In Section~5, we give evaluations of double
integrals arising in quantum field theory.

\subsection{Hypergeometric series}

We may also obtain a simple hypergeometric series for $c_{3,0}$
from the Clausen product formula~\cite[p.\ 178]{agm} in the form
\begin{equation}\frac{4}{\pi^2}{\bf K}^2(\sin(\alpha/2))={}_3{\rm F}_2
\left(\begin{array}{cccc}\frac12,\frac12,\frac12\\
1,1\end{array}\bigg|\sin^2\alpha\right)
\label{Clausen}
\end{equation}
which is valid for $\pi/2\ge\alpha\ge0$. Setting $\alpha=\pi/6$,
so that $\sin(\alpha)=1/2$,
we recast~(\ref{c30}) as
\begin{equation}
c_{3,0}=
\frac{\sqrt3\pi^3}{8}\sum_{n=0}^\infty\frac{{2n\choose n}^3}{2^{8n}}
= \frac{\sqrt3\pi^3}{8}\,{}_3{\rm F}_2
\left(\begin{array}{cccc}\frac12,\frac12,\frac12\\
1,1\end{array}\bigg|\frac 14\right).
\label{c30hyp}
\end{equation}
Moreover, we conjectured the compact formula
\begin{equation}
c_{3,2}=
\frac{\sqrt3\pi^3}{288}\,\sum_{n=0}^\infty\frac{{2n\choose n}^3}{2^{8n}}
\frac{1}{(n+1)^2}=\frac{\sqrt3\pi^3}{288}\,{}_3{\rm F}_2
\left(\begin{array}{cccc}\frac12,\frac12,\frac12\\
2,2\end{array}\bigg|\frac 14\right)
\label{c32hyp}
\end{equation}
as an alternative to~(\ref{c32pslq}). A proof was provided by
\emph{Maple}, which evaluates this sum in terms
of $K_3$ and $E_3$.
Our evaluation of the latter then shows that~(\ref{c32hyp})
follows from~(\ref{c32}).

\subsection{Integrals of elliptic integrals}

Setting $a=b=1$, $w=2\tan\theta$ and
$c=2\sin\alpha$ in~(\ref{novee3}), we obtain
from~(\ref{AandB})
the evaluation
\[\int_0^{\pi/2}
\frac{{\bf K}(\sin\theta)}
{\cos\theta\sqrt{\tan^2\theta+\sin^2\alpha}}
\,{\rm d}\theta=\frac{B(\sin\alpha)}{2\sin\alpha}\]
and hence, by trigonometric simplification, we prove the identity
\begin{equation}
\int_0^{\pi/2}\frac{{\bf K}(\sin\theta)}
{\sqrt{1-\cos^2\alpha\cos^2\theta}}\,{\rm d}\theta
={\bf K}(\sin(\alpha/2))\,{\bf K}(\cos(\alpha/2))
\label{Honolulu}
\end{equation}
which we had conjectured empirically from interpolating known results, using
\emph{Maple}'s {\tt MinimalPolynomial}.
At $\alpha=\pi/2$, identity~(\ref{Honolulu})
reduces to the evaluation
\[\int_0^1\frac{{\bf K}(k)}{\sqrt{1-k^2}}\,{\rm d}k=
{\bf K}^2\left(\frac{1}{\sqrt2}\right)=\frac{\Gamma^4\left(\frac14\right)}
{16\pi}\]
given in~\cite[p.\ 188]{agm}. At $\alpha=\pi/6$, we obtain
\begin{equation}
2\int_0^{\pi/2}\frac{{\bf K}(\sin\theta)}
{\sqrt{1+3\sin^2\theta}}\,{\rm d}\theta=K_3K_3^\prime\,.
\label{jonc30}
\end{equation}

We did not find identity~(\ref{Honolulu}) in the literature. However,
John Zucker remarked to us that the left-hand side
may be transformed to a double integral using~(\ref{Kk}). If we then
exchange the order of integration and set
$C=\cos^2\alpha$ and $S=\sin^2\phi$ in the evaluation
\[\int_0^{\pi/2}
\frac{1}{\sqrt{(1-C\cos^2\theta)(1-S\sin^2\theta)}}\,{\rm d}\theta
={\bf K}\left(\sqrt{C+S-C S}\right)\]
given in~\cite{MK}, we obtain an identity first derived by
Glasser~\cite{Gla}, namely
\begin{equation}
\int_0^{\pi/2}\,{\bf K}\left(\sqrt{1-\sin^2\alpha\cos^2\phi}\right)
\,{\rm d}\phi
={\bf K}(\sin(\alpha/2))\,{\bf K}(\cos(\alpha/2))\,,
\label{alpha}
\end{equation}
which was re-derived by Joyce and Zucker
and recorded in~\cite[Eq.\ 2.3.5]{expm2}.

\subsection{Sum rule}

Using the analysis of $c_{3,0}$ above and in~\cite{bbbc07}
we obtain a sum rule
\begin{equation}
\sum_{n=0}^\infty\frac{{2n\choose n}^3}{2^{8n}}\left(
\frac83\log2-\sum_{k=1}^n\frac{1}{k(2k-1)}
-\frac{\pi}{\sqrt3}\right)
=0\,.
\label{DJBdisE3}
\end{equation}
In Section~5 we conjecture an integral counterpart to this sum rule.

\section{Four Bessel functions}

We may construct $S_4$ by folding one instance of $S_3$, in~(\ref{ess3}),
with the discontinuity of another, to obtain
\begin{equation}
S_4(a,b,c,w)=\int_{a+b}^\infty 2 u D_3(a,b,u)S_3(u,c,w)\,{\rm d}u
\label{ess4int}
\end{equation}
from which we obtain an evaluation of the \emph{on-shell} value
\begin{equation}
s_{4,1}:=S_4(1,1,1,{\rm i})
=\int_2^\infty\frac{4\,{\rm arctanh}
\left(\sqrt{\frac{u-2}{u+2}}\right)}
{u(u^2-4)}\,{\rm d}u
=\int_0^1\frac{2y\log(y)}{y^4-1}\,{\rm d}y=\frac{\pi^2}{16}
\label{ess4onshell}
\end{equation}
by the substitution $u=y+1/y$.

\subsection{The odd moments $s_{4,2k+1}$}

By differentiation of
\[S_4(1,1,1,{\rm i}x)=\int_2^\infty\frac{4\,{\rm arctanh}
\left(\sqrt{\frac{(u-1)^2-x^2}{(u+1)^2-x^2}}\right)}
{\sqrt{(u^2-4)((u-1)^2-x^2)((u+1)^2-x^2)}}\,{\rm d}u\]
we evaluate
\[s_{4,3}=
\left.\left(\frac{1}{x}+\frac{{\rm d}}{{\rm d}x}\right)
\frac{{\rm d}S_4(1,1,1,{\rm i}x)}{{\rm d}x}\right|_{x=1}
=\frac{\pi^2}{64}\]
and are then able to solve the recursion relation~\cite{bsalvy,ouvry}
for $s_{4,2k+1}$ by the \emph{closed form}
\begin{equation}
s_{4,2k+1}=\frac{\pi^2}{16}\left(\frac{k!}{4^k}\right)^2b_k
\label{ess4k}
\end{equation}
with integers
\begin{equation}
b_k=\sum_{j=0}^k{k\choose j}^2{2k-2j\choose k-j}{2j\choose j}\,.
\label{A2895}
\end{equation}

Integer sequence~(\ref{A2895}) begins
\begin{equation}
1,\,4,\,28,\,256,\,2716,\,31504,\,387136,\,4951552,\,65218204,\,
878536624,\,12046924528
\label{A2895s}
\end{equation}
and is recorded\footnote{See {\tt
http://www.research.att.com/\~{ }njas/sequences/A002895}~.}
as entry A2895 of the on-line version of~\cite{EIS},
which gives the recursion
\begin{equation}
k^3b_k-2(2k-1)(5k^2-5k+2)b_{k-1}+64(k-1)^3b_{k-2}=0
\label{diamond}
\end{equation}
and the generating function
\begin{equation}
I_0^4(2t)=\sum_{k=0}^\infty b_k\left(\frac{t^k}{k!}\right)^2\,.
\label{A2895g}
\end{equation}
We have verified that recursion~(\ref{diamond}) reproduces the recursion
for~(\ref{ess4k}), which has the same form as for the odd
moments in~(\ref{salvy4}).
We note that in~\cite{crystals,matrices} the integers $b_k$
were related to enumeration of paths in three-dimensional diamond lattices.
They also appear in a study~\cite[Eq.\ 6.10]{walks} of lattice
magnetic walks. {}From the generating function $I_0^4$ we see that
they result from the constrained sum
\[b_k=\sum_{p+q+r+s=k}\left(\frac{k!}{p!\,q!\,r!\,s!}\right)^2\,.\]
The economical recursion in~(\ref{diamond}) will be used in Section~5.

\subsection{Dispersion relation}

We adapt the dispersive method of~\cite{Barton,2loop,Sabry}
to two spacetime dimensions and
take the discontinuity of~(\ref{ess4int})
across the cut with
branch point $w^2=-(a+b+c)^2$, obtaining the dispersion relation
\begin{equation}S_4(a,b,c,w)
=\int_{a+b+c}^\infty\frac{2v\,D_4(a,b,c,v)}{v^2+w^2}\,{\rm d}v
\label{ess4}
\end{equation}
with a discontinuity given by a complete elliptic integral {\bf K}
of the first kind,
and hence by an arithmetic-geometric mean~\cite{agmhist,agm}, namely
\begin{eqnarray}
D_4(a,b,c,d)&=&\int_{a+b}^{d-c}2u\,D_3(a,b,u)D_3(u,c,d)\,{\rm d}u
\label{Dalitz}\\
&=&\frac{2{\bf K}\left(\frac{Q(a,b,c,-d)}{Q(a,b,c,d)}\right)}{Q(a,b,c,d)}
=\frac{\pi}{{\rm AGM}\left(Q(a,b,c,d),\,4\sqrt{a b c d}\right)}
\label{dee4}
\end{eqnarray}
where
\[Q(a,b,c,d):=\sqrt{(a+b+c+d)(a+b-c-d)(a-b+c-d)(a-b-c+d)}\]
is completely symmetric in its 4 arguments.

In physical terms, $D_4(a,b,c,d)$ gives the volume of phase
space for the decay of a particle of mass $d>a+b+c$ into 3 particles
with masses $a$, $b$ and $c$ in two spacetime dimensions. In 4
spacetime dimensions, one would obtain an incomplete elliptic
integral for the area of the Dalitz plot,
in the generic mass case; in two spacetime dimensions we obtain
a simple arithmetic-geometric mean for the length of a Dalitz
line~\cite{davdel}.

Using this elliptic representation for $S_4(1,1,1,0)=V_3(1,1,1)=c_{3,1}$
we obtain an evaluation of the integral
\begin{equation}
\int_{0}^\frac13 D(y)\,{\rm d}y=c_{3,1}=\frac34L_{-3}(2)
\label{evalDv3}
\end{equation}
with $D(y):=2D_4(1,1,1,1/y)/y$ given by
\begin{equation}
D(y)=\frac{4y\,{\bf K}\left(\sqrt{\frac{(1-3y)(1+y)^3}{(1+3y)(1-y)^3}}\right)}
{\sqrt{(1+3y)(1-y)^3}}=\frac{3\sqrt{3}\pi y}{2}\,
{\rm HeunG}\left(-8,-2;1,1,1,1; 1-9y^2\right)
\label{dee}
\end{equation}
where the \emph{general Heun} function, {\tt HeunG},
satisfies \emph{Heun's differential equation}
as specified in \emph{Maple}, see~\cite{heun}.
Similarly,
\[\int_{0}^\frac13 D(y)y^2\,{\rm d}y=\frac14\,c_{3,3}
=\frac14\,L_{-3}(2)-\frac16\,.\]

In Section 5 we shall show that the {\tt HeunG} representation
of the elliptic integral~(\ref{dee}), from quantum field theory,
may be related to the hexagonal lattice sequence~(\ref{A2893})
of crystal theory. For the present, we note
that~(\ref{ess4onshell}) and~(\ref{ess4})
provide the evaluation
\begin{equation}
\int_{0}^\frac13\frac{D(y)}{1-y^2}\,{\rm d}y=s_{4,1}=\frac{\pi^2}{16}\,.
\label{evalDs4}
\end{equation}

\subsection{The odd moments $c_{4,2k+1}$}

For $V_4$ we have two representations.
First, from the elementary {\tt arctanh} function in~(\ref{ess3})
we may compute
\begin{equation}
V_4(a,b,c,d)=\int_0^\infty w S_3(a,b,w)S_3(c,d,w)\,{\rm d}w
\label{vee4}
\end{equation}
and easily evaluate
\[c_{4,1}=\int_0^\infty\frac{4\,{\rm arctanh}^2\left(\frac{w}{\sqrt{w^2+4}}
\right)}{w(w^2+4)}\,{\rm d}w
=\int_0^1\frac{4y\log^2(y)}{1-y^4}\,{\rm d}y=\frac78\zeta(3)\]
by the substitution $w=1/y-y$.
Similarly, by differentiation of~(\ref{vee4}), we obtain
\[c_{4,3}=\frac{7}{32}\zeta(3)-\frac{3}{16}\,.\]
In general, all the odd moments $c_{4,2k+1}$ are given
by rational linear combinations of $\zeta(3)$ and unity,
as shown in~\cite{bbbc07,bsalvy,ouvry}. Because of the mixing of $\zeta(3)$
with unity, we were unable to write a closed form for their
rational coefficients in $c_{4,2k+1}$.

The alternative folding, using~(\ref{dee4}), is
\begin{eqnarray*}
V_4(a,b,c,d)&=&\int_0^\infty w S_4(a,b,c,w)S_2(d,w)\,{\rm d}w\\
&=&\int_{a+b+c}^\infty2v\,D_4(a,b,c,v)V_2(v,d)\,{\rm d}v,
\end{eqnarray*}
which yields a novel formula for $\zeta(3)$, namely
\begin{equation}
\int_{0}^\frac13\frac{D(y)\log(y)}{y^2-1}\,{\rm d}y=c_{4,1}=\frac78\zeta(3)
\label{evalDv4}
\end{equation}
by the substitution $v=1/y$.

\subsection{The even moment $c_{4,0}$}

The analogue to~(\ref{vee4}), for even moments, is
\begin{equation}
\overline{V}_4(a,b,c,d)=2\pi\int_0^\infty
\overline{S}_3(a,b,w)\overline{S}_3(c,d,w)\,{\rm d}w
\label{novee4}
\end{equation}
with a product of elliptic integrals coming from~(\ref{noess3}).
Setting $c=a$ and $d=b$, we obtain the intriguing case
\begin{equation}
\int_0^\infty K_0^2(a t)K_0^2(b t)\,{\rm d}t
=2\pi\int_0^\infty
\frac{{\bf K}^2\left(\sqrt{\frac{w^2+(a-b)^2}{w^2+(a+b)^2}}\right)}
{w^2+(a+b)^2}\,{\rm d}w
\label{kab}
\end{equation}
where the square of {\bf K} may be replaced by a ${}_3{\rm F}_2$
series~\cite{bailey1935} for only part of the range of the integration,
because of the restricted validity of the Clausen product~(\ref{Clausen}).
In particular it was not at first clear how to evaluate the integral in
\begin{equation}
c_{4,0}=\pi\int_0^{\pi/2}{\bf K}^2(\sin\theta)\,{\rm d}\theta
\label{c40int}
\end{equation}
obtained by setting $a=b=1$ and $w=2\tan\theta$.

The key to unlock
this puzzle was provided by the trigonometric series
\begin{equation}{\bf K}(\sin\theta)=
\sum_{n=0}^\infty \gamma_n\sin((4n+1)\theta)
\label{djbfgt}
\end{equation}
with coefficients
\begin{equation}
\gamma_n:=\left(\frac{\Gamma\left(n+\frac12\right)}{\Gamma(n+1)}\right)^2
=\frac{4}{4n+1}+{\rm O}\left(\frac1{n^3}\right)
\label{gamma}
\end{equation}
given by F.G.\ Tricomi~\cite{tricomi}
and recorded in~\cite[Section~13.8, Eqn.~(8)]{erdelyi}.
The identity~(\ref{djbfgt}) is valid for $\pi/2>\theta>0$.
Thus, the integral in~(\ref{c40int}) is easily transformed to
a hypergeometric sum in
\begin{equation}
c_{4,0}=\frac{\pi^2}{4}\sum_{n=0}^\infty\gamma_n^2
=\frac{\pi^4}{4}\sum_{n=0}^\infty\frac{{2n\choose n}^4}{2^{8n}}
=\frac{\pi^4}{4}\;{}_4{\rm F}_3
\left(\begin{array}{ccccc}\frac12,\frac12,\frac12,\frac12\\
1,1,1\end{array}\bigg|1\right)
\label{c40hyp}
\end{equation}
by using the orthogonality relation
\[\int_0^{\pi/2}\sin((4m+1)\theta)\,\sin((4n+1)\theta)\,{\rm d}\theta=
\frac{\pi}{4}\delta_{m,n}\,.\]

It is instructive to split $c_{4,0}={\mathcal A}_1+{\mathcal A}_2$
into the contributions
\[{\mathcal A}_1=\pi\int_0^{\pi/4}{\bf K}^2(\sin\theta)\,{\rm d}\theta\,,\quad
{\mathcal A}_2=\pi\int_0^{\pi/4}{\bf K}^2(\cos\theta)\,{\rm d}\theta\,,\]
and use the Clausen product~(\ref{Clausen}) for the former.
The result is ${\mathcal A}_1=\frac14c_{4,0}$ which proves that
${\mathcal A}_2/{\mathcal A}_1=3$,
as we had noticed numerically.
A more direct derivation
of this proven factor of 3 would probably be enlightening.

We also note that ${\mathcal A}_1=\frac14c_{4,0}$
may be obtained by setting $\phi=0$ in the double series
\[\pi\int_{\phi/2}^{\pi/4}{\bf K}^2(\sin\theta)\,{\rm d}\theta
=\frac{\pi^3}{8}\cos\phi\sum_{n=0}^{\infty}
\frac{{2n\choose n}^3}{2^{6n}}\;{}_2{\rm F}_1
\left(\begin{array}{ccc}\frac12,\frac12-n\\
\frac 32\end{array}\bigg|\cos^2\phi\right)\]
which is valid $\pi/2\ge\phi\ge0$.

\subsection{Sum rule}

Combining~(\ref{c40hyp}) with a more complicated sum
in~\cite[Eq.\ 3-5]{bbbc07},
we get the discrete sum rule
\begin{equation}
\sum_{n=0}^\infty\frac{{2n\choose n}^4}{2^{8n}}
\left\{8\left(-\log2+\sum_{k=1}^n\frac{1}{2k(2k-1)}\right)^2
-\sum_{k=1}^n\frac{4k-1}{2k^2(2k-1)^2}-\frac{\pi^2}{3}\right\}
=0
\label{DJBdisE4}
\end{equation}
with one more central binomial coefficient
and one more power of $\pi/\sqrt{3}$ than~(\ref{DJBdisE3}).

\subsection{The even moment $c_{4,2}$}

Integer relation algorithms (see~\cite{pslq,ema,2level}
for extended discussion) led us to conjecture that
\begin{equation}
c_{4,2}=\frac{\pi^2}{256}
\sum_{n=0}^\infty \gamma_n^2
\left(\frac{12}{n+1}-\frac{3}{(n+1)^2}-8\right)
\label{c42hyp}
\end{equation}
which we shall now prove.

As before, we use Bessel's differential equation and operate
on our master formula, in this case~(\ref{novee4}),
before setting parameters to unity, obtaining
\[c_{4,2}=\pi\int_0^{\pi/2}
\left(\frac{\cos4\theta-1}{16}
+\frac{4\cot\theta+\sin4\theta}{32}\,
\frac{{\rm d}}{{\rm d}\theta}\right)\,{\bf K}^2(\sin\theta)
\,{\rm d}\theta\,.\]
Then, using Tricomi's expansion~(\ref{djbfgt}),
we easily reduce
\[\pi\int_0^{\pi/2}\,{\bf K}^2(\sin\theta)\,\cos4\theta\,
{\rm d}\theta=\frac{\pi^2}{16}
\sum_{n=0}^\infty \gamma_n^2
\left(2-\frac{1}{n+1}\right)^2\]
to single sums of the form in~(\ref{c42hyp}).
The term involving $\sin4\theta$ gives a multiple
of the same sum, using integration by parts.

However, the term involving $\cot\theta$ is more demanding.
Integrating it by parts, we conclude that~(\ref{c42hyp}) is
equivalent to the evaluation
\begin{equation}
\mathcal{B}:=\int_0^{\pi/2}\frac{4\pi\,{\bf K}^2(\sin\theta)-\pi^3}
{\sin^2\theta}\,{\rm d}\theta=
\frac{\pi^2}{4}\sum_{n=0}^\infty \gamma_n^2
\frac{4n+3}{(n+1)^2}
\label{c42both}
\end{equation}
which we now prove by using a subtracted form for the differential
of~(\ref{djbfgt}).

Defining $\delta_n:=4-(4n+1)\gamma_n$, we prove by induction that
\begin{equation}
\sum_{m=0}^{n-1}\delta_m=4n-4n^2\,\gamma_n=1-\sum_{m=n}^\infty \delta_m\,.
\label{Dsum}
\end{equation}
Then from this we derive a subtracted series for
\begin{equation}
\frac{{\rm d}{\bf K}(\sin\theta)}{{\rm d}\theta}-
\frac{\sin^2\theta}{\cos\theta}=
\sum_{m=1}^\infty \delta_m(\cos\theta-\cos((4m+1)\theta))
\label{djbsub}
\end{equation}
by subtracting the trigonometric series~\cite{fts}
\begin{equation}
4\sum_{m=0}^\infty
\frac{\sin((4m+1)\theta)}{4m+1}=\frac{\pi}{2}+\log(\sec\theta+\tan\theta)
\label{sec}
\end{equation}
from~(\ref{djbfgt}) and then differentiating, to obtain
\[\frac{{\rm d}{\bf K}(\sin\theta)}{{\rm d}\theta}-\sec\theta=
-\sum_{m=1}^\infty \delta_m\cos((4m+1)\theta)\,.\]
Adding $\cos\theta$ to each side and using $\sum_{m=0}^\infty \delta_m=1$,
as a consequence of~(\ref{Dsum}), we obtain~(\ref{djbsub}).

By combining~(\ref{djbfgt}) and~(\ref{djbsub}), we obtain
the double series
\[\frac{\mathcal{B}}{2\pi^2}-\gamma_0=\sum_{m,n\ge0}\delta_m\left(
M_{0,n}-M_{m,n}\right)\gamma_n\]
using the triangular array $M$ with entries
\[M_{m,n}:=\frac{4}{\pi}\int_0^{\pi/2}
\cot\theta\,\cos((4m+1)\theta)\,\sin((4n+1)\theta)\,{\rm d}\theta\]
for non-negative integers $m$ and $n$. These entries are very simple:
$M_{m,n}=2$, for $m<n$, $M_{m,n}=0$, for $m>n$, and $M_{n,n}=1$,
resulting in
\[\frac{\mathcal{B}}{2\pi^2}=\gamma_0+
\sum_{m=0}^\infty \delta_m\left(
-\gamma_m-\gamma_0+2\sum_{n=0}^m \gamma_n\right)
=\sum_{n=0}^\infty \gamma_n \varepsilon_n\]
with the $\gamma_0$ terms cancelling, since $\sum_{m=0}^\infty \delta_m=1$.
Here,
\[\varepsilon_n:=-\delta_n+2\sum_{m=n}^\infty \delta_m
=(8n^2+4n+1)\gamma_n-2(4n+1)\]
appears by a change of the order of summation and is easily evaluated,
by using~(\ref{Dsum}).

Looking back to what needs to be proved, in~(\ref{c42both}),
we see that we now need to establish the vanishing of
\[\sum_{n=0}^\infty
\gamma_n\left(\frac{4n+3}{8(n+1)^2}\gamma_n-\varepsilon_n\right)=0\,.\]
This is achieved by taking the $N\to\infty$ limit
of the explicit evaluation
\begin{equation}
\sum_{n=0}^{N-1}
\gamma_n\left(\frac{4n+3}{8(n+1)^2}\gamma_n-\varepsilon_n\right)
=2N^2\, \gamma_N \delta_N=O\left(\frac 1N \right)
\label{nogosp}
\end{equation}
of a truncated sum, which is easily proven by induction.
We note that the {\tt Gosper} algorithm~\cite{petkovsek} in \emph{Maple}
failed to evaluate~(\ref{nogosp}), as written,
since $\delta_n$ and $\varepsilon_n$ mix binomial and polynomial terms.
If one separates these by hand then our compact result
is verified.

\subsection{Further evaluation of integrals}

We also succeeded in separating $\mathcal{B}=\mathcal{B}_1+\mathcal{B}_2$
into the contributions
\begin{eqnarray}
\mathcal{B}_1&:=&\int_0^{\pi/4}\frac{4\pi\,{\bf K}^2(\sin\theta)-\pi^3}
{\sin^2\theta}\,{\rm d}\theta=
-2\pi^2+\mathcal{H}
+\frac{\pi^2}{4}\sum_{n=0}^\infty \gamma_n^2\,
\frac{2n+1}{n+1}\quad\quad
\label{c42lo}\\
\mathcal{B}_2&:=&\int_0^{\pi/4}\frac{4\pi\,{\bf K}^2(\cos\theta)-\pi^3}
{\cos^2\theta}\,{\rm d}\theta=
-2\pi^2-\mathcal{H}
+\frac{3\pi^2}{4}\sum_{n=0}^\infty \gamma_n^2\,
\frac{2n+1}{n+1}\quad\quad
\label{c42hi}
\end{eqnarray}
with a familiar factor of 3 multiplying the ${}_4{\rm F}_3$
series in the latter and a new constant from the singular value
$k_1=\sin(\pi/4)$, namely
\begin{eqnarray*}
\mathcal{H}&:=&\pi^3
\left(1-\,{}_3{\rm F}_2\left(\begin{array}{cccc}-\frac12,\frac12,\frac12\\
1,1\end{array}\bigg|1\right)\right)\\
&=&\pi^3-
\frac{\Gamma^4\left(\frac14\right)+16\,\Gamma^4\left(\frac34\right)}{8}\,.
\end{eqnarray*}
The $\pi^3$ terms in $\mathcal{H}$ and the integrand
of~(\ref{c42hi}) match, since
$\int_0^{\pi/4}\sec^2\theta\,{\rm d}\theta=1$.

\subsection{Further evaluations of sums}

Evaluations~(\ref{c42lo}) and~(\ref{c42hi})
were discovered using PSLQ. To prove~(\ref{c42lo}),
we may again use a Clausen product, with its first term subtracted.
We then encounter the integrals
\[2\int_0^{\pi/4}\frac{\sin^{2n}2\theta}{\sin^2\theta}\,
{\rm d}\theta=B\left(n-\frac12,\frac12\right)+\frac{2}{2n-1}\]
with $n>0$. Here $B$ is Euler's Beta function and yields
yet another ${}_4{\rm F}_3$ series, which we eliminate by
using the $N\to\infty$ limit of the summation
\[\sum_{n=0}^{N-1}\gamma_n^2
\left(\frac{2}{1-2n}-\frac{1}{n+1}\right)
=\frac{16N^3}{2N-1}\gamma_N^2=8+{\rm O}\left(\frac 1N \right)\]
which was also proven by induction.
To prove~(\ref{c42hi}), we then subtract~(\ref{c42lo})
from~(\ref{c42both}) and use
the $N\to\infty$ limit of the summation
\[\sum_{n=0}^{N-1}\gamma_n^2
\left(8-\frac{8}{n+1}+\frac{1}{(n+1)^2}\right)
=16N^2\gamma_N^2=16+{\rm O}\left(\frac 1N \right)\]
which was proven in the same manner.

Thus one may undo the
explicit evaluations of parts of~(\ref{c42hi}),
in terms of $\pi$ and $\Gamma$ values, and instead write
\[\pi\int_0^{\pi/4}
\frac{{\bf K}^2(\cos \theta )}{\cos^2\theta}\,{\rm d}\theta
=\frac{\pi^3}{8}\sum_{n=0}^\infty\frac{{2n\choose n}^3}{2^{6n}}
\frac{2n+1}{n+1}
+\frac{\pi^4}{32}\sum_{n=0}^\infty\frac{{2n\choose n}^4}{2^{8n}}
\frac{4n^2+10n+5}{(n+1)^2}\]
in terms of undigested hypergeometric series.

More productively, we may use the explicit summations so far achieved
to reduce
\begin{eqnarray}
\frac{16c_{4,0}-64c_{4,2}}{3\pi^4}-\frac{4}{\pi^2}=
{}_4{\rm F}_3\left(\begin{array}{cccc}\frac12,\frac12,\frac12,\frac12\\2,
1,1 \end{array}\bigg|1\right)
\label{c4single}
\end{eqnarray}
to a single ${}_4{\rm F}_3$ series. Comparing this last result
with~(\ref{c40hyp}), we conclude
that all moments $c_{4,2k}$ can be expressed in terms $\pi$
and a pair of contiguous ${}_4{\rm F}_3$ series.
An equivalent hypergeometric expression is
\begin{equation}
c_{4,2}=\frac{\pi^4}{64}\left\{4\;{}_4{\rm F}_3
\left(\begin{array}{ccccc}\frac12,\frac12,\frac12,\frac12\\
1,1,1\end{array}\bigg|1\right)-3\;{}_4{\rm F}_3
\left(\begin{array}{ccccc}\frac12,\frac12,\frac12,\frac12\\
2,1,1\end{array}\bigg|1\right)\right\}-\frac{3\pi^2}{16}\,.
\label{c42hyp2}
\end{equation}
For comparison we repeat
\begin{equation}
c_{4,0}=\frac{\pi^4}{4}\;{}_4{\rm F}_3
\left(\begin{array}{ccccc}\frac12,\frac12,\frac12,\frac12\\
1,1,1\end{array}\bigg|1\right)\,.
\label{c40hyp2}
\end{equation}

\subsection{Relation to Meijer's G-function}

A generalization of~(\ref{kab}) is derivable
in terms of the \emph{Meijer-G function}~\cite[Vol.\ 3]{prud}.
For example, we have
\begin{equation}
I(a,b,k):=\int_0^{\infty}t^k K_0^2(a t)K_0^2(b t)\,{\rm d}t
=\frac{\pi}{8a^{k+1}}G^{33}_{44}\left(\frac{b^2}{a^2}\left|
\begin{array}{cccc}
\frac{1-k}{2}&\frac{1-k}{2}&\frac{1-k}{2}&\frac{1}{2}\\
0&0&0&-\frac{k}{2}
\end{array}\right.\right)
\label{Iabk}
\end{equation}
which may be proven by making two copies of the representation
\[t^{\mu}K_0^2(a t)=\frac{\sqrt{\pi}}{2a^{\mu}}G^{30}_{13}
\left(a^2t^2\left|\begin{array}{ccc}
&\frac{\mu+1}2&\\
\frac{\mu}2&\frac{\mu}2&\frac{\mu}2
\end{array}\right.\right)\]
and integrating them with weight $t$ to obtain
\[I(a,b,\mu+\nu+1)=\frac{\pi}{4a^{\mu}b^{\nu}}\int_0^{\infty}t\,G^{30}_{13}
\left(a^2t^2\left|\begin{array}{ccc}
&\frac{\mu+1}2&\\
\frac{\mu}2&\frac{\mu}2&\frac{\mu}2
\end{array}\right.\right)G^{30}_{13}\left(b^2t^2\left|\begin{array}{ccc}
&\frac{\nu+1}2&\\
\frac{\nu}2&\frac{\nu}2&\frac{\nu}2
\end{array}\right.\right)\,{\rm d}t\,.\]
Then we use Meijer's result for the integral of the product of two
G-functions to obtain
\[I(a,b,\mu+\nu+1)=\frac{\pi}{8a^{\mu+2}b^{\nu}}G^{33}_{44}
\left(\frac{b^2}{a^2}\left|\begin{array}{cccc}
-\frac \mu 2&-\frac \mu 2&- \frac \mu 2&\frac{\nu+1}2\\
\frac \nu 2& \frac \nu 2&\frac \nu 2&-\frac{\mu+1}2
\end{array}\right.\right)\,.\]
The apparent freedom in the choice of parameters $\mu$ and $\nu$
is demystified by formula 8.2.1.15 in~\cite[Vol.\ 3]{prud},
which shows that multiplication by a power of the argument
of a G-function is equivalent to adding a constant
to all its parameters, as in the example~(\ref{Bmeijer}).
We resolve this redundancy by setting $\mu=k-1$ and $\nu=0$
and hence prove~(\ref{Iabk}).
Similarly, we obtained
\begin{equation}
\int_0^{\infty}t^k K_0(a t)K_0^2(b t)\,{\rm d}t
=\frac{2^{k-2}\sqrt{\pi}}{a^{k+1}}\,
G^{32}_{33}\left(\frac{4b^2}{a^2}\left|
\begin{array}{ccc}
\frac{1-k}{2}&\frac{1-k}{2}&\frac{1}{2}\\
0&0&0
\end{array}\right.\right)\,.
\label{gee3}
\end{equation}

In~\cite{bbbc07} the special cases of~(\ref{Iabk}) and~(\ref{gee3})
with $a=b=1$ were studied numerically,
using Adamchik's algorithm~\cite{adam}. This algorithm
converges quickly in the case of $c_{3,k}$, obtained from~(\ref{gee3})
with an argument of 4. But in \emph{Maple} it is painfully slow
in the case of $c_{4,k}$, obtained from~(\ref{Iabk}) with unit argument.
Numerical evaluation of our new hypergeometric
results~(\ref{c42hyp2}) and~(\ref{c40hyp2}) goes far faster, with \emph{Maple}.

\subsection{Another continued fraction}

Again we may derive from~\cite{bbbc07} that
\begin{equation}
8\,\frac{c_{4,2}}{c_{4,0}}=
\cfrac{1^6}{e(1)-\cfrac{3^6}{\ddots-\cfrac{(2N-1)^6}{e(N)-\dots}}}
\label{contfracc4e}
\end{equation}
where $e(N):=N(20N^2 + 3)$ and the
ratio $8\,c_{4,2}/c_{4,0}$ may be made explicit
from~(\ref{c42hyp2},\ref{c40hyp2}).

\subsection{The even moment $s_{4,0}$}

The odd moment $s_{4,1}$ relates directly to quantum field theory;
it is the two-loop on-shell equal-mass sunshine diagram
in two spacetime dimensions. No such meaning attaches to $s_{4,0}$;
it is hard to think of a physical application for this moment.
However, we found a rather pretty formula for it, which we record as
\begin{equation}
s_{4,0}:=\int_0^\infty I_0(t)K_0^3(t)\,{\rm d}t
=\int_0^{\pi/2}{\bf K}(\sin\theta)\,{\bf K}(\cos\theta)\,{\rm d}\theta\,.
\label{s40}
\end{equation}
This amusing twist of the integral~(\ref{c40int}) for $c_{4,0}$ follows
from Nicholson's integral representation~\cite[13.72, Eq.\ 3]{watson}
of the product
\[I_0(t)K_0(t)=\frac{2}{\pi}\int_0^{\pi/2}K_0(2t\sin\alpha)\,{\rm d}\alpha\,.\]
Substituting $c=2\sin\alpha$ in~(\ref{AandB}) and using the
appropriate reduction~(\ref{Bform}) to a product of {\bf K} values,
we obtain
\[\int_0^\infty I_0(t)K_0^3(t)\,{\rm d}t=
\int_0^{\pi/2}\frac{B(\sin\alpha)}{2\sin\alpha}\,{\rm d}\alpha=
\int_0^{\pi/2}{\bf K}(\sin(\alpha/2))\,{\bf K}(\cos(\alpha/2))
\,{\rm d}\alpha\,.\]
Setting $\alpha=2\theta$, we prove~(\ref{s40}). Then \emph{Pari-GP} gives,
in a tenth of a second, 64 digits of
\[s_{4,0}=
6.997563016680632359556757826853096005697754284353362908336255807\ldots\]
A corresponding twist of the sum~(\ref{c40hyp}) for $c_{4,0}$
comes from Tricomi's expansion~(\ref{djbfgt}) and
\[\int_0^{\pi/2}\sin((4m+1)\theta)\cos((4n+1)\theta)\,{\rm d}\theta
=\frac{1}{4m+4n+2}\,.\]
This yields the double sum
\[s_{4,0}=\sum_{m,n\ge0}\frac{\gamma_m\gamma_n}{4m+4n+2}=
\sum_{n=0}^\infty\gamma^2_n\left(\lambda_n-\frac{1}{8n+2}\right)\]
with $\gamma_n$ in~(\ref{gamma}) and
\[\lambda_n:=\frac{2}{\gamma_n}\sum_{m=0}^n\frac{\gamma_m}{4m+4n+2}
=1+\sum_{m=1}^n\frac{8m}{16m^2-1}\]
where the latter form was proven by \emph{Maple}. Hence we obtain
\begin{equation}
s_{4,0}:=\int_0^\infty I_0(t)K_0^3(t)\,{\rm d}t
=\frac{\pi^2}{2}\sum_{n=0}^\infty\frac{{2n\choose n}^4}{2^{8n}}
\left(\frac{1}{4n+1}+\sum_{k=0}^{2n-1}\frac{2}{2k+1}\right)\,.
\label{s40sumalt}
\end{equation}
By way of comparison, we note the simpler evaluation
\begin{equation}
\int_0^\infty I_0^2(t)K_0^2(t)\,{\rm d}t
=\sum_{k=0}^\infty{2k\choose k}\frac{c_{2,2k}}{(2^k{}k!)^2}
=\frac{\pi^2}{4}\sum_{k=0}^\infty\frac{{2k\choose k}^4}{2^{8k}}
=\frac{1}{\pi^2}\int_0^\infty K_0^4(t)\,{\rm d}t
\label{t40}
\end{equation}
from the closed form for $c_{2,2k}$ in~(\ref{closed}) and the expansion
\begin{equation}
I_0^2(t)
=\sum_{k=0}^\infty{2k\choose k}\left(\frac{t^k}{2^k{}k!}\right)^2\,.
\label{I02}
\end{equation}

\subsection{Tabular summary}

In Table 1 we recapitulate the key discoveries for the moments
$c_{n,k}:=\int_0^\infty t^k K_0^n(t)\,{\rm d}t$ with $n=3,4$.
The results for the even moments $c_{3,2k}$ and $c_{4,2k}$ are new.
The table may extended by using the recursions~(\ref{salvy3})
and~{(\ref{salvy4}).

\begin{table}\label{table1}
$$\begin{array}{|rcl|}
\hline
c_{3,0}&=&\displaystyle\frac{3\,\Gamma^6\left(\frac13\right)}{32\pi 2^{2/3}}
\;=\;\frac{\sqrt{3}\pi^3}{8}\;
{}_3{\rm F}_2\left(\begin{array}{c}\frac12,\frac12,\frac12\\1,1\end{array}
\bigg|\frac14\right){\rule{0pt}{5.0ex}}\\[10pt]
c_{3,1}&=&\displaystyle\frac{3}{4}L_{-3}(2)\\[10pt]
c_{3,2}&=&\displaystyle\frac{\Gamma^6\left(\frac{1}{3}\right)}{96\pi 2^{2/3}}
-\frac{4\pi^5 2^{2/3}}{9\,\Gamma^6\left(\frac{1}{3}\right)}
\;=\;\frac{\sqrt{3}\pi^3}{288}\;
{}_3{\rm F}_2\left(\begin{array}{c}\frac12,\frac12,\frac12\\2,2\end{array}
\bigg|\frac14\right)\\[10pt]
c_{3,3}&=&\displaystyle L_{-3}(2)-\frac{2}{3}\\[10pt]
\hline
c_{4,0}&=&\displaystyle\frac{\pi^4}{4}
\sum_{n=0}^\infty\frac{{2n\choose n}^4}{2^{8n}}
\;=\;\frac{\pi^4}{4}\;{}_4{\rm F}_3\left(
\begin{array}{c}\frac12,\frac12,\frac12,\frac12\\1,1,1\end{array}
\bigg|1\right){\rule{0pt}{5.0ex}}\\[10pt]
c_{4,1}&=&\displaystyle\frac{7}{8}\zeta(3)\\[10pt]
c_{4,2}&=&\displaystyle\frac{\pi^4}{16}\;{}_4{\rm F}_3\left(
\begin{array}{c}\frac12,\frac12,\frac12,\frac12\\1,1,1\end{array}
\bigg|1\right)
-\frac{3\pi^4}{64}\;{}_4{\rm F}_3\left(
\begin{array}{c}\frac12,\frac12,\frac12,\frac12\\2,1,1\end{array}
\bigg|1\right)
\displaystyle-\frac{3\pi^2}{16}\\[10pt]
c_{4,3}&=&\displaystyle\frac{7}{32}\zeta(3)-\frac{3}{16}\\[10pt]
\hline
\end{array}$$
\caption{Evaluations for $c_{n,k}$ with $n=3,4$.}
\end{table}

\section{Five Bessel functions}

Little is known for certain about integrals involving 5 Bessel functions.
However, there are some remarkable conjectures arising from studies
in quantum field theory~\cite{Lap205,Lap21}.

\subsection{Conjectural evaluations of Feynman diagrams}

In~\cite{Lap205}, Stefano Laporta developed an impressive technique
for numerical evaluation of the coefficients of the Laurent expansion
in $\varepsilon$ of Feynman diagrams in $D=4-2\varepsilon$ spacetime
dimensions. Here we are concerned with just one of the many diagrams
that he considered, namely the dimensionally regularized
3-loop sunrise diagram with 4 internal lines:
\[{\mathcal S}_5(w^2,D):=
\int\int\int\left.\frac{{\rm d}^D p_1\,{\rm d}^D p_2\,{\rm d}^D p_3}
{N(p_1)N(p_2)N(p_3)N(q-p_1-p_2-p_3)}\right|_{q\cdot q=w^2}\]
where $N(p):=p\cdot p+1$ is the inverse propagator of a scalar
particle with unit mass and momentum $p$.
In the on-shell case, the Laurent
expansion found by Laporta has the form
\[\frac{{\mathcal S}_5(-1,4-2\varepsilon)}
{\left(\pi^{2-\varepsilon}\Gamma(1+\varepsilon)\right)^3}
=\frac{2}{\varepsilon^3}
+\frac{22}{3\varepsilon^2}
+\frac{577}{36\varepsilon}+S_{205}+{\rm O}(\varepsilon)\]
with a numerical value $S_{205}\approx21.92956264368$, for the finite part,
given in equation~(205) of~\cite{Lap205}.
Subsequently, in equation~(21) of~\cite{Lap21},
this constant was conjecturally related to products of elliptic
integrals of the first and second kind, with a numerical
check to 1200 decimal places.

In a talk\footnote{Zentrum f\"ur interdisziplin\"are
Forschung in Bielefeld, 14th of June, 2007. Displays available from {\tt
http://www.physik.uni-bielefeld.de/igs/schools/ZiF2007/Broadhurst.pdf}
leading to 200,000 decimal places for $c_{5,1}$ and $c_{5,3}$ in {\tt
http://paftp.open.ac.uk/pub/staff\_ftp/dbroadhu/newconst/V5AB.txt}~.}
entitled ``Reciprocal PSLQ and the Tiny Nome of Bologna",
Broadhurst observed that Laporta's conjecture may be written
rather intriguingly using the constant
\begin{equation}
C:=\frac{\pi}{16}\left(1-\frac{1}{\sqrt{5}}\right)
\left(1+2\sum_{n=1}^\infty\exp\left(-n^2\pi\sqrt{15}\right)\right)^4
\label{CDef}
\end{equation}
and its \emph{reciprocal} $1/C$, in terms of which he wrote
Laporta's conjecture as
\begin{equation}
S_{205}\stackconj{1}
\frac{6191}{216}-\frac{\pi^2}{3}\left(4C+\frac{7}{40C}\right)\,.
\label{Lap}
\end{equation}

Broadhurst further conjectured that the odd moments
\[s_{5,2k+1}:=\int_0^\infty t^{2k+1}I_0(t)K_0^4(t)\,{\rm d}t\]
are linear combinations of $\pi^2C$ and $\pi^2/C$, with rational coefficients,
and in particular that
\begin{eqnarray}
\frac{s_{5,1}}{\pi^2}&\stackconj{2}&C
\label{s51}\\
\frac{s_{5,3}}{\pi^2}&\stackconj{3}&
\left(\frac{2}{15}\right)^2\left(13C-\frac{1}{10C}\right)
\label{s53}\\
\frac{s_{5,5}}{\pi^2}&\stackconj{4}&
\left(\frac{4}{15}\right)^3\left(43C-\frac{19}{40C}\right)
\label{s55}
\end{eqnarray}
with higher moments obtained by a recursion of the
form~(\ref{littlecrec}) with polynomials $p_{5,i}$
given in~(\ref{crec56}). We have checked these 3 conjectures
to 1200 decimal places.

In the course of this work, we discovered that the moments
\[t_{5,2k+1}:=\int_0^\infty t^{2k+1}I_0^2(t)K_0^3(t)\,{\rm d}t\]
follow an uncannily similar pattern. If $q_k$ and $r_k$ are the rational
numbers that give $s_{5,2k+1}/\pi^2$, conjecturally, as $q_k C-r_k/C$,
then we found that
$q_k C+r_k/C$ gives the value of $2t_{5,2k+1}/(\sqrt{15}\pi)$.
We checked 60,000 decimal places of the resultant evaluations
\begin{eqnarray}
\frac{2t_{5,1}}{\sqrt{15}\pi}&\stacktick{5}&C
\label{t51}\\
\frac{2t_{5,3}}{\sqrt{15}\pi}&\stacktick{6}&
\left(\frac{2}{15}\right)^2\left(13C+\frac{1}{10C}\right)
\label{t53}\\
\frac{2t_{5,5}}{\sqrt{15}\pi}&\stacktick{7}&
\left(\frac{4}{15}\right)^3\left(43C+\frac{19}{40C}\right)
\label{t55}
\end{eqnarray}
for which we eventually found a proof, presented in subsection~\ref{proof}.

Finally, by doubling one of the masses in the Feynman diagram
corresponding to $t_{5,1}$, we arrived at the conjectural
evaluation
\begin{equation}
\int_0^\infty t\,I_0^2(t)K_0^2(t)K_0(2t)\,{\rm d}t
\stackconj{8}\frac{1}{12}\,K_3K_3^\prime
\label{DJB3a}
\end{equation}
which has been checked to 1200 decimal places.

\paragraph{Notation.} In the 8 evaluations~(\ref{Lap}) to~(\ref{DJB3a})
we have used the device $\stackconj{n}$ or $\stacktick{n}$
to distinguish the cases that remain unproven from the 3
cases in~(\ref{t51}) to~(\ref{t55}), which we were eventually
able to prove. Some of the labels $n=1\ldots8$
will recur, as we give equivalent forms of these
conjectured or proven evaluations.

\subsection{The odd moments $t_{5,2k+1}$}

Evaluations~(\ref{t51},\ref{t53},\ref{t55}) were easy to
check to high precision, thanks to our closed form~(\ref{ess4k})
for the odd moments $s_{4,2k+1}$. By expanding one of the
functions
\[I_0(t)=\sum_{n=0}^\infty\left(\frac{t^n}{2^n{}n!}\right)^2\]
in the integrand, $t^{2k+1}I_0^2(t)K_0^3(t)$, of $t_{5,2k+1}$,
we obtain a rapidly converging sum in
\begin{equation}
t_{5,2k+1}=4^{k-2}\pi^2\sum_{n=k}^\infty b_n
\left(\frac{n!}{8^n(n-k)!}\right)^2
\label{60k}
\end{equation}
in terms of the diamond lattice integers~(\ref{A2895}).
To relate $t_{5,1}$ to a product of complete elliptic integrals
we use Jacobi's identity
\begin{equation}
\sqrt{\frac{2\,{\bf K}(k)}{\pi}}=\theta_3(q):=
\sum_{n=-\infty}^\infty q^{n^2}
\label{theta3}
\end{equation}
with a \emph{nome} related to $k$ by
$q=\exp(-\pi\,{\bf K}^\prime(k)/{\bf K}(k))$.
Specializing to the \emph{singular value}~\cite{agm}
\[k_{15}=\frac{(2-\sqrt3)(\sqrt5-\sqrt3)(3-\sqrt5)}{8\sqrt2}\]
with the ``tiny nome"
$q_{15}:=\exp(-\pi\sqrt{15})\approx0.000005197$,
we obtain from~(\ref{CDef})
\[C=\frac{\sqrt5-1}{4\sqrt5\pi}\,K_{15}^2=\frac{1}{2\sqrt{15}\pi}\,
K_{15}K_{5/3}\]
with $K_{15}:={\bf K}(k_{15})$ and
$K_{5/3}:={\bf K}(k_{5/3})$, where
\[k_{5/3}=\frac{(2-\sqrt3)(\sqrt5+\sqrt3)(3+\sqrt5)}{8\sqrt2}\]
yields the larger nome $q_{5/3}:=\exp(-\pi\sqrt{5/3})=q_{15}^{1/3}$.
Thus evaluation~(\ref{t51}) amounts to
\begin{equation}t_{5,1}
=\frac{\pi^2}{16}\sum_{n=0}^\infty\frac{b_n}{64^n}
\stacktick{5}\frac14\,K_{15}K_{5/3}
\label{t51sum}
\end{equation}
with a summand $b_n/64^n={\rm O}(n^{-3/2}/4^n)$, from~(\ref{A2895}),
giving rapid convergence.
By taking $10^5$ terms, we checked~(\ref{t51}) to 60,000 decimal places,
using the recursion~(\ref{diamond}) for the diamond lattice sequence
$b_n$. Our closed form in~(\ref{t51sum}) resulted
from paying diligent attention to a footnote in~\cite[p.\ 121]{Lap21},
which led us, eventually, to discover and prove the connection
between quantum field theory and these diamond lattice integers.

We remark that our
evaluations of $s_{5,2k+1}$ and $t_{5,2k+1}$ may be expressed in terms
of $\Gamma$ values, using the corresponding evaluation of $K_{15}^2$
in~\cite{ema}, as was remarked by Laporta after Broadhurst's talk
(see footnote~3). For example, we may re-write the conjectural evaluation
for $s_{5,3}$ in~(\ref{s53}) as
\begin{equation}
\frac{\sqrt{5}}{2}\int_0^\infty t^3I_0(t)K_0^4(t)\,{\rm d}t
\stackconj{3}\frac{
13\,\Gamma\left(\frac{1}{15}\right)
\Gamma\left(\frac{2}{15}\right)
\Gamma\left(\frac{4}{15}\right)
\Gamma\left(\frac{8}{15}\right)}{30^3}-
\frac{\Gamma\left(\frac{7}{15}\right)
\Gamma\left(\frac{11}{15}\right)
\Gamma\left(\frac{13}{15}\right)
\Gamma\left(\frac{14}{15}\right)}{15}
\label{Gamma3}
\end{equation}
which contains all 8 values of $\Gamma(n/15)$ with $n\in[1,14]$
and coprime to 15. Then the counterpart for $t_{5,3}$ in~(\ref{t53})
may be written as
\begin{equation}
\frac{\pi^3}{4\sqrt3}\sum_{n=1}^\infty\frac{n^2b_n}{64^n}
\stacktick{6}\frac{
13\,\Gamma\left(\frac{1}{15}\right)
\Gamma\left(\frac{2}{15}\right)
\Gamma\left(\frac{4}{15}\right)
\Gamma\left(\frac{8}{15}\right)}{30^3}+
\frac{\Gamma\left(\frac{7}{15}\right)
\Gamma\left(\frac{11}{15}\right)
\Gamma\left(\frac{13}{15}\right)
\Gamma\left(\frac{14}{15}\right)}{15}
\label{Gamma6}
\end{equation}
by the remarkable sign change discovered in our present work
and the relation to diamond lattice numbers in~(\ref{60k}).

\subsection{Double integrals}

The moments $c_{5,1}$, $s_{5,1}$ and $t_{5,1}$
are easily expressible as double integrals of elementary functions.

For the 4-loop \emph{vacuum}~\cite{Lap21} diagram
\[V_5(a,b,c,d,e):=
\int_0^\infty t\,K_0(a t)K_0(b t)K_0(c t)K_0(d t)K_0(e t)\,{\rm d}t\]
in two spacetime dimensions, we obtain the double integral
\[V_5(a,b,c,d,e)=
\int_0^\infty\int_0^\infty x y S_3(a,b,x)D_3(x,y,{\rm i}c)S_3(d,e,y)
\,{\rm d}x\,{\rm d}y\]
by grouping the internal lines with masses $a$ and $b$ to give
a total momentum with norm $x^2$ and those with masses $d$ and $e$
to give a total momentum with norm $y^2$. Then the coupling term
\[\frac{1}{\pi}\int_0^\pi\frac{{\rm d}\theta}
{x^2+2 x y\cos\theta+y^2+c^2}
=\frac{1}{\sqrt{(x+y)^2+c^2}\sqrt{(x-y)^2+c^2}}
=D_3(x,y,{\rm i}c)\]
comes from an angular average in two Euclidean dimensions.
Setting the five masses to unity, we obtain
\begin{equation}
c_{5,1}=
\int_0^\infty\int_0^\infty\frac{4\,{\rm arcsinh}(x/2)\,{\rm arcsinh}(y/2)
\,{\rm d}x\,{\rm d}y}{\sqrt{(4+x^2)(4+y^2)(1+(x+y)^2)(1+(x-y)^2)}}
\label{c51alt}
\end{equation}
where we have converted the {\tt arctanh} function of~(\ref{ess3})
to a more convenient {\tt arcsinh} function, in the equal-mass case.

Similarly, the 3-loop \emph{sunrise}~\cite{groote} diagram
\[S_5(a,b,c,d,z):=
\int_0^\infty t\,K_0(a t)K_0(b t)K_0(c t)K_0(d t)J_0(z t)\,{\rm d}t\]
yields the double integral
\[S_5(a,b,c,d,z)=\int_0^\infty\int_0^\infty x y
S_3(a,b,x)D_3(x,y,{\rm i}c)D_3(y,z,{\rm i}d)
\,{\rm d}x\,{\rm d}y\]
by cutting one line in $V_5$ and setting the norm of its Euclidean
momentum to $z^2$. Setting the 4 masses to unity and analytically
continuing to the on-shell point $z^2=-1$, we obtain
\begin{equation}
s_{5,1}=\int_0^\infty\int_0^\infty\frac{2\,{\rm arcsinh}(x/2)
\,{\rm d}x\,{\rm d}y}{\sqrt{(4+x^2)(4+y^2)(1+(x+y)^2)(1+(x-y)^2)}}
\stackconj{2}\frac{\pi}{2\sqrt{15}}\,K_{15}K_{5/3}
\label{s51alt}
\end{equation}
whose conjectural evaluation is given by~(\ref{s51}).

We then define
\[T_5(w,a,b,c,z):=
\int_0^\infty t\,J_0(w t)K_0(a t)K_0(b t)K_0(c t)J_0(z t)\,{\rm d}t\]
as the angular average of the diagram obtained by cutting
two lines in $V_5$ and setting the norms of their momenta
to $w^2$ and $z^2$. Hence we obtain the double integral
\[T_5(w,a,b,c,z)=\int_0^\infty\int_0^\infty x y
D_3(w,x,{\rm i}a)D_3(x,y,{\rm i}b)D_3(y,z,{\rm i}c)
\,{\rm d}x\,{\rm d}y\]
which leads to
\begin{equation}
t_{5,1}=\int_0^\infty\int_0^\infty\frac{
\,{\rm d}x\,{\rm d}y}{\sqrt{(4+x^2)(4+y^2)(1+(x+y)^2)(1+(x-y)^2)}}
\stacktick{5}\frac{1}{4}\,K_{15}K_{5/3}
\label{conj5}
\end{equation}
whose evaluation is given by~(\ref{t51}).
If we multiply~(\ref{conj5}) by 4, we recover the double-integral
discovery reported by Laporta in Eqs. 17, 18 and 19 of~\cite{Lap21}.

Finally, by doubling one of the internal masses, we obtain
\[\int_0^\infty t\,I_0^2(t)K_0^2(t)K_0(2t)\,{\rm d}t=
\int_0^\infty\int_0^\infty\frac{
\,{\rm d}x\,{\rm d}y}{\sqrt{(4+x^2)(4+y^2)(4+(x+y)^2)(4+(x-y)^2)}}\]
and hence the conjectural evaluation
\begin{equation}
\int_0^\infty\int_0^\infty\frac{
\,{\rm d}x\,{\rm d}y}{\sqrt{(1+x^2)(1+y^2)(1+(x+y)^2)(1+(x-y)^2)}}
\stackconj{8}\frac13\,K_3K_3^\prime
\label{conj8}
\end{equation}
after rescaling $x$ and $y$ by a factor of 2 and
invoking~(\ref{DJB3a}).

Evaluation~(\ref{conj8}) resonates with the proven
evaluations~(\ref{DJB3b}) and~(\ref{DJB3c}) in Section 3, where
we found that removing $(1+(x-y)^2)$ from the square root in~(\ref{conj8})
doubles the value of the integral and that
removing $(1+(x+y)^2)$ multiplies it by 4.

We were unable find a transformation of variables
for the double integrals in~(\ref{conj5}) and~(\ref{conj8})
that suggested their evaluations as products at the singular values
$k_{15}$ and $k_{3}$, respectively. In the next 3 subsections
we show how to express~(\ref{conj5}) as a single integral of
a complete elliptic integral, in 3 rather different ways.

As observed in a footnote in~\cite[p.\ 120]{Lap21}, entry 3.1.5.16
in~\cite[Vol.\ 1]{prud} is intriguing:
for real parameters, $k_1$ and $k_2$ with $k_1^2+k_2^2<1$ one has
\begin{equation}
\int_0^{\pi/2}\int_0^{\pi/2} \frac{{\rm d}\theta\,{\rm d}\phi}
{\sqrt{1-k_1^2\sin^2\theta-k_2^2\sin^2\phi}}
=\frac{2\,{\bf K}(\alpha)\,{\bf K}^\prime(\beta)}{1+\sqrt{1-k_2^2}}
\label{prod2}
\end{equation}
where
\[\alpha:=\frac{\sqrt{1-k_1^2}-\sqrt{1-k_1^2-k_2^2}}{1+\sqrt{1-k_2^2}}\,
\mbox{ and }\,
\beta:=\frac{\sqrt{1-k_1^2}+\sqrt{1-k_1^2-k_2^2}}{1+\sqrt{1-k_2^2}}.\]
Perhaps there are implications for~(\ref{conj8})
from this general form.
As discussed in~\cite{barzan, Gla2}, one can establish
a more recondite counterpart for
\[\int_0^{\pi/2}\int_0^{\pi/2}
\sqrt{1-k_1^2\sin^2\theta-k_2^2\sin^2\phi}
\,\,{\rm d}\theta\,{\rm d}\phi\,.\]

\subsection{Single integrals from polar coordinates}

We may recast evaluations~(\ref{conj5}) and~(\ref{conj8})
by transforming the more general double integral
\[J(c):=
\int_0^\infty\int_0^\infty\frac{{\rm d}x\,{\rm d}y}
{\sqrt{(c+x^2)(c+y^2)(1+(x+y)^2)(1+(x-y)^2)}}\]
to polar coordinates.
With $x=r\cos\theta$ and $y=r\sin\theta$,
the product of the first two factors in the square root gives
a term linear in $w:=\cos4\theta$, as does the remaining product.
An angular integral of the form
\[\int_{-1}^1\frac{{\rm d}w}{\sqrt{(1-w^2)(2A^2-1-w)(2B^2-1+w)}}=
\frac{1}{A B}\,{\bf K}\left(\frac{\sqrt{A^2+B^2-1}}{A B}\right)\]
results, with $A=1+2c/r^2$ and $B=1+1/r^2$. Transforming to
$z=r^2/(1+r^2)$, we obtain the single integral
\begin{equation}
J(c)=\int_0^1{\bf K}\left(\frac{z\sqrt{1+4c(1-z)(z+c(1-z))}}
{z+2c(1-z)}\right)\frac{{\rm d}z}{z+2c(1-z)}\,.
\label{jay}
\end{equation}

Setting $c=4$, we transform evaluation~(\ref{conj5}) to
\begin{equation}
\int_0^1{\bf K}\left(\frac{z\sqrt{(13-12z)(5-4z)}}{8-7z}\right)
\frac{{\rm d}z}{8-7z}
\stacktick{5}\frac14\,K_{15}K_{5/3}\,.
\label{conj5z}
\end{equation}

Setting $c=1$, we transform conjecture~(\ref{conj8}) to
\begin{equation}
\int_0^1{\bf K}\left(\frac{z\sqrt{5-4z}}{2-z}\right)\frac{{\rm d}z}{2-z}
\stackconj{8}\frac 13\,K_3K_{1/3}
\label{conj8z}
\end{equation}
since, by definition of a singular value, $K_{1/3}=K_3^\prime$.
A hypergeometric version of~(\ref{conj8z}) may be obtained by
writing its left-hand side as the ornate triple sum
\[\frac{\pi}{4}\sum_{n=0}^\infty\left(\frac{(2n)!}{2^{2n}n!}\right)^3
\sum _{m=0}^n2^{2m}\,\frac{{}_2{\rm F}_1
\left(\begin{array}{ccc}2n+1,2n+1\\
2n+2+m\end{array}\bigg|\frac 12\right)}{(n-m)!(2n+1+m)!}
\stackconj{8}\frac{\pi^2}{4\sqrt3}\;{}_3{\rm F}_2
\left(\begin{array}{cccc}\frac 12, \frac 12,\frac 12\\
1,1\end{array}\bigg|\frac 14\right)~.
\end{equation*}

We computed the single integrals in~(\ref{conj5z}) and~(\ref{conj8z})
using \emph{Pari-GP}, which provides
an efficient {\tt agm} procedure, for evaluating the complete elliptic
integral {\bf K}, and an efficient {\tt intnum} procedure, for the
numerical quadrature. In each case, we confirmed the evaluation
to 1200 decimal places but were none the wiser as to its origin.

\subsection{Single integrals over discontinuities}

Seeking illumination, we turned to integrals over the
elliptic integral~(\ref{dee4}) in the discontinuity
$D_4$, coming from the Dalitz-plot integration in~(\ref{Dalitz}).

For the 4-loop vacuum diagram $V_5$ we may fold $D_4$ from~(\ref{dee4})
with $V_3$ from~(\ref{vee3}), to obtain
\begin{equation}
V_5(a,b,c,d,e)=\int_{a+b+c}^\infty 2 v D_4(a,b,c,v)V_3(v,d,e)\,{\rm d}v\,.
\label{vee5}
\end{equation}
With unit masses and $y=1/v$, this gives the moments
\begin{eqnarray}
c_{5,1}&=&\int_{0}^\frac13\frac{D(y)L(y)}{\sqrt{1-4y^2}}\,{\rm d}y
\label{c51}\\
c_{5,3}&=&\int_{0}^\frac13\frac{y^2D(y)}{(1-4y^2)^2}\left(
\frac{4(1-2y^2+4y^4)L(y)}{\sqrt{1-4y^2}}
+2-8y^2+8(1-y^2)\log(y)\right)\,{\rm d}y\quad\quad
\label{c53}
\end{eqnarray}
with a complete elliptic integral in $D(y)$, from~(\ref{dee}), and
a dilogarithmic function
\begin{eqnarray}
L(y)&:=&\frac12
{\rm Li}_2\left(\frac{\sqrt{1-4y^2}-1}{\sqrt{1-4y^2}+1}\right)
-\frac12
{\rm Li}_2\left(\frac{\sqrt{1-4y^2}+1}{\sqrt{1-4y^2}-1}\right)
\label{Ly2}\\
&=&-{\rm Li}_2\left(\frac{1-\sqrt{1-4y^2}}{2}\right)
+\frac12\log^2\left(\frac{1-\sqrt{1-4y^2}}{2}\right)
-\log^2(y)+\frac{\pi^2}{12}
\label{Ly1}
\end{eqnarray}
with~(\ref{Ly2}) coming from~(\ref{vee3}) and the reduction
to a single convenient ${\rm Li}_2$
value in~(\ref{Ly1}) obtained by transformations
in~\cite[A.2.1]{lewin} and noted in~\cite[Eq.\ 3.21]{davtau0}.
Further differentiations yield an even lengthier integrand for $c_{5,5}$.
However, we expected the odd moments $c_{5,2k+1}$ with $k>1$ to be
expressible as Q-linear combinations of $c_{5,3}$, $c_{5,1}$
and unity. Hence we lazily used PSLQ to arrive at our ninth conjecture
\begin{equation}
c_{5,5}\stackconj{9}\frac{76}{15}\,c_{5,3}-\frac{16}{45}\,c_{5,1}
+\frac{8}{15}
\label{c55}
\end{equation}
with a final rational term originating, presumably, from the
analytically trivial input
\begin{equation}
\int_0^\infty t^5K_0(t)K_1^4(t)\,{\rm d}t
=\lim_{t\to0}\frac{t^5K_1^5(t)}{5}=\frac15
\label{trivial}
\end{equation}
to the richer (and more challenging) recursions considered in
the talk of footnote~3, which dealt with integrals of products
of powers of $t$, $K_0(t)$
and the derivative $K_0^\prime(t)=-K_1(t)$. {}From~(\ref{c51}),~(\ref{c53})
and the eminently believable conjecture~(\ref{c55}),
higher odd moments are obtainable from~(\ref{littlecrec})
and~(\ref{crec56}). Unfortunately we lack a more explicit evaluation of
the single integrals for $c_{5,1}$ and $c_{5,3}$.

For the 3-loop sunrise diagram, the corresponding folding is
\begin{equation}
S_5(a,b,c,d,w)
=\int_{a+b+c}^\infty2v\,D_4(a,b,c,v)S_3(v,d,w)\,{\rm d}v
\label{ess5b}
\end{equation}
from which we obtain the on-shell value
\begin{equation}
s_{5,1}=\int_{0}^\frac13\frac{2D(y)}{\sqrt{1-4y^2}}
\,{\rm arctanh}\left(\sqrt{\frac{1-2y}{1+2y}}\right)\,{\rm d}y
\stackconj{2}\frac{\pi}{2\sqrt{15}}\,K_{15}K_{5/3}
\label{Ds51}
\end{equation}
and, by use of Bessel's equation, the more complicated integral
\begin{equation}
s_{5,3}=\int_{0}^\frac13\frac{4y^2D(y)}{(1-4y^2)^2}
\,\left(\frac{2(1-2y^2+4y^4)\,
{\rm arctanh}\left(\sqrt{\frac{1-2y}{1+2y}}\right)}{\sqrt{1-4y^2}}
-1+y^2\right)\,{\rm d}y
\label{Ds53}
\end{equation}
for the next odd moment, conjecturally given by~(\ref{s53}).
Then PSLQ gives
\begin{equation}
s_{5,5}\stackconj{2,3,4}\frac{76}{15}\,s_{5,3}-\frac{16}{45}\,s_{5,1}
\label{s55rec}
\end{equation}
which presumably results from a partial integration in the even
richer (and even more challenging) recursions for integrals of
products of powers of $t$, $I_0(t)$, $I_1(t)$, $K_0(t)$ and $K_1(t)$.
We used \emph{Pari-GP} to evaluate the dispersive single integrals
for the first 3 odd moments $s_{5,2k+1}$, with $k=0,1,2$,
to 1200 decimal places, and hence checked the
conjectured evaluations~(\ref{s51}),~(\ref{s53})
and~(\ref{s55}) and their consequent integer relation~(\ref{s55rec}),
to this high precision.

Similarly, the folding
\begin{equation}
T_5(u,a,b,c,w)
=\int_{a+b+c}^\infty2v\,D_4(a,b,c,v)D_3(u,{\rm i}v,w)\,{\rm d}v
\label{tee5b}
\end{equation}
gives, with $y=1/v$, the on-shell, unit-mass result
\begin{equation}
t_{5,1}=\int_{0}^\frac13\frac{D(y)}{\sqrt{1-4y^2}}\,{\rm d}y
\stacktick{5}\frac{1}{4}\,K_{15}K_{5/3}
\label{Dt51}
\end{equation}
and, by use of Bessel's equation,
\begin{equation}
t_{5,3}=\int_{0}^\frac13\frac{4y^2(1-2y^2+4y^4)D(y)}{(1-4y^2)^{5/2}}\,{\rm d}y
\label{Dt53}
\end{equation}
for the next odd moment, given by~(\ref{t53}).
Then PSLQ gives
\begin{equation}
t_{5,5}\stacktick{7}\frac{76}{15}\,t_{5,3}-\frac{16}{45}\,t_{5,1}
\label{t55rec}
\end{equation}
which was checked to 60,000 decimal places, in the more convenient sum
\begin{equation}
\sum_{n=0}^\infty\frac{b_n}{64^n}\left(16n^2(n-1)^2-\frac{76}{15}\,4n^2
+\frac{16}{45}\right)\stacktick{7}0
\label{t55sum}
\end{equation}
over the diamond lattice integers in~(\ref{60k}).

By setting $u=w={\rm i}$, $a=b=1$ and $c=2$ in~(\ref{tee5b})
and transforming to $y=4/v$, we obtain the representation
\begin{equation}
\int_0^1\frac{4y\,{\bf K}\left(
\frac{2+y}{2-y}\sqrt{\frac{1-y}{1+y}}\right)}
{\sqrt{(2-y)^3(2+y)(1+y)}}\,{\rm d}y
\stackconj{8}\frac13K_3K_{1/3}
\label{conj8d}
\end{equation}
as the dispersive counterpart to~(\ref{conj8z}),
but still have no better idea as to why the third singular
value occurs.

\subsection{Relations between elliptic integrals}

We found a third way of stating our conjectures in terms of integrals
over ${\bf K}$. This arose from the double-integral representation
\begin{equation}
s_{5,1}=\int_0^\infty\int_0^\infty
\frac{2\,{\rm arcsinh}(x/2)\,{\rm d}x\,{\rm d}y}
{\sqrt{(4+x^2)(4+y^2)R(x,y)}}
\stackconj{2}\frac{\pi}{2\sqrt{15}}\,K_{15}K_{5/3}
\label{Rform}
\end{equation}
with the product $(1+(x+y)^2)(1+(x-y)^2)$ in~(\ref{s51alt})
replaced by the non-factorizable term
\[R(x,y)=\left|\frac{(x+{\rm i})^2+4+y^2}{2}\right|^2
=x^2+\left(\frac{3+x^2+y^2}{2}\right)^2.\]

We obtained~(\ref{Rform})
by introducing a Feynman parameter, $z$,
to combine two unit-mass propagators, with momenta
$q-p$ and $r-q$, in the integral
\[\frac{1}{A B}
=\int_0^1\frac{{\rm d}z}{(z A+(1-z)B)^2}\]
with $A:=N(q-p)$ and $B:=N(r-q)$. Here,
$p$ is the combined momentum of the other two
internal lines and $r$
is the external momentum of the 3-loop sunrise diagram.
Integration over the two-dimensional vector $q$
leaves an integral over $z$ with a denominator
$z(1-z)(r-p)\cdot(r-p)+1$ that is symmetric
about the mid-point $z=\frac12$.
Integrating over the angle between $p$ and $r$,
setting $p\cdot p=x^2$ and analytically continuing
to the on-shell point $r\cdot r=-1$, we obtain~(\ref{Rform})
by making the transformation $z(1-z)=1/(4+y^2)$
which maps $z=0$ to $y=\infty$ and the mid-point
$z=\frac12$ to $y=0$.

This method provided an analytical advance, since it proved the
equality $E_1(u)=E_2(u)$ between the elliptic integrals
\begin{eqnarray}
E_1(u)&:=&\frac12\int_0^\infty\frac{{\rm d}v}
{\sqrt{v(4+v)(1+2u+2v+(u-v)^2)}}
\label{E1}\\
E_2(u)&:=&\int_0^\infty\frac{{\rm d}v}
{\sqrt{v(4+v)(4u+(3+u+v)^2)}}
\label{E2}
\end{eqnarray}
where we have transformed to $u=x^2$ and $v=y^2$. When \emph{Maple} was
asked to evaluate $E_1$ and $E_2$ it printed different
expressions, each containing an incomplete elliptic integral
\[F(\sin\phi,k):= \int_0^\phi\frac{1}
{\sqrt{1-k^2\sin^2\theta}}\,{\rm d}\theta\]
with a relation between $k$ and $\phi$ which we reduced to the form
\begin{equation}
k=\frac{\sqrt{1-2\cos\phi}}{(1-\cos\phi)\sin\phi}\,.
\label{kphi}
\end{equation}
The incomplete integral was eliminated by computing $(2E_1+E_2)/3$,
for which \emph{Maple} gave an expression involving only
the complete elliptic integral ${\bf K}(k)$. By this means, we proved that
\begin{equation}
F(\sin\phi,k)=\frac23\,{\bf K}(k)
\label{E1E2}
\end{equation}
whenever $k$ and $\phi$ are related by condition~(\ref{kphi}).
Moreover, by using~\cite[17.4.13]{AandS}, we proved that this condition
also gives the evaluation
\begin{equation}
F(1-\cos\phi,k)=\frac13\,{\bf K}(k)
\label{E1E2bis}
\end{equation}
which we shall use in the next subsection.

Our result for $E(x):=E_1(x^2)=E_2(x^2)$ is
\begin{equation}
E(x)=\sqrt{\frac{\sin(2\alpha(x))}{2x}}\frac{{\bf K}(\sin\alpha(x))}{3}
=\frac{\pi}{3\sqrt3}\,{\rm HeunG}(9,3;1,1,1,1;-x^2)
\label{eee}
\end{equation}
with
\begin{equation}
\alpha(x)=\frac{3\arctan(x)-\arctan(x/3)}{2}\,.
\label{alpha-x}
\end{equation}
This then gives
\begin{eqnarray}
s_{5,1}=\int_{0}^\infty\frac{2E(x)}{\sqrt{4+x^2}}
\,{\rm arcsinh}\left(\frac{x}{2}\right)\,{\rm d}x\
&\stackconj{2}&\frac{\pi}{2\sqrt{15}}\,K_{15}K_{5/3}
\label{Es51}\\
t_{5,1}=\int_{0}^\infty\frac{E(x)}{\sqrt{4+x^2}}\,{\rm d}x\
&\stacktick{5}&\frac{1}{4}\,K_{15}K_{5/3}
\label{Et51}
\end{eqnarray}
as the non-dispersive counterparts
to the integrals~(\ref{Ds51}) and~(\ref{Dt51})
over the different complete elliptic integral in $D(y)$,
with a closely related {\tt HeunG} function in~(\ref{dee}).

We note that $E(x)$ contains a factor of $1/3$ from
evaluating $(2E_1+E_2)/3$. This ensures that we correctly reproduce
\[s_{3,1}=E(0)=\frac{\pi}{3\sqrt{3}}=L_{-3}(1)\]
using $\alpha(x)=4x/3+{\rm O}(x^3)$. The relationship
\begin{equation}
\tan(2\alpha(x))=\frac{8x}{3-6x^2-x^4}
\label{tan2}
\end{equation}
gives $E(x)=\log(x^2)/x^2+{\rm O}(\log(x)/x^4)$, as expected at large
momentum.

\subsection{Expansions near singularities}

The {\tt HeunG} forms for $D(y)$ and $E(x)$, in~(\ref{dee})
and~(\ref{eee}), yield
expansions near the regular points $y=\frac13$ and $x=0$, respectively.
However, these regular expansions were not needed in our numerical
integrations, since the {\tt agm} procedure of
\emph{Pari-GP} is highly efficient for the evaluation of
${\bf K}(k)$ when $k$ is not close to the singular point at $k=1$.
What we really need, and eventually found, are the expansions
\begin{eqnarray}
\frac{D(y)}{3y}&=&
-\sum_{k=0}^\infty\left(h_k+a_k\log(y^2)\right)y^{2k}
\label{Dseries}\\
x^2E(x)&=&-\sum_{k=0}^\infty\frac{h_k-a_k\log(x^2)}{(-x^2)^k}
\label{Eseries}
\end{eqnarray}
that isolate the logarithmic singularities as $y\to0$
and $x\to\infty$, respectively. Here $a_k$ is the hexagonal lattice
integer of~(\ref{A2893}) and $h_k$ is determined by the
differential equation for $D(y)$, or equivalently $E(x)$, which
yields the recursion
\begin{equation}
(k+1)^2h_{k+1}-(10k^2+10k+3)h_k+9k^2h_{k-1}=
-2(k+1)a_{k+1}+10(2k+1)a_k-18ka_{k-1}
\label{hrec}
\end{equation}
with a starting value $h_0=0$. We note that~(\ref{Dseries})
converges for $0<y<\frac13$ and~(\ref{Eseries}) converges for $x>3$.
We were alerted to role of the hexagonal lattice integers
by the regular expansion
\begin{equation}
E(x)=\frac{\pi}{3\sqrt3}\sum_{k=0}^\infty a_k\left(\frac{-x^2}{9}\right)^k
\label{Esmallx}
\end{equation}
which is valid for $|x|<1$, since $a_k={\rm O}(9^k/k)$ for large $k$.

We were able to solve the recursion~(\ref{hrec})
in closed form, by considering the moment
\[T_4(u,a,b,v):=
\int_0^\infty t\,J_0(u t)K_0(a t)K_0(b t)J_0(v t)\,{\rm d}t
=\int_{a+b}^\infty 2w\,D_3(a,b,w)D_3(u,v,{\rm i}w)\,{\rm d}w\]
which yields, in general, an incomplete elliptic integral
\begin{equation}
T_4(u,a,b,v)=
\int_{(a+b)^2}^\infty\frac{{\rm d}x}
{\sqrt{(x-(a+b)^2)(x-(a-b)^2)(x+(u+v)^2)(x+(u-v)^2)}}
\label{tee4}
\end{equation}
by transforming to $x=w^2$. In the special case with $a=1$, $b=1/y>1$
and $u=v={\rm i}$, \emph{Maple} gave the evaluation
\[T_4({\rm i},1,1/y,{\rm i})
=\frac{2y^2F(1-\cos\phi,k)}{\sqrt{(1+3y)(1-y)^3}}\]
with
\begin{equation}
\phi=\arccos\left(\frac{2y}{1+y}\right)\,,\quad\quad
k=\sqrt{\frac{(1-3y)(1+y)^3}{(1+3y)(1-y)^3}}\,.
\label{kphiD}
\end{equation}
This relation between $k$ and $\phi$ is precisely as in~(\ref{kphi}).
Hence, using~(\ref{E1E2bis}), we obtain
\begin{equation}
\frac{y\,D(y)}{6}=T_4({\rm i},1,1/y,{\rm i}):=
\int_0^\infty t\,I_0^2(t)K_0(t)K_0(t/y)\,{\rm d}t
\label{by6}
\end{equation}
since $D(y)$ in~(\ref{dee}) contains the complete elliptic integral
${\bf K}(k)$
with $k$ given in~(\ref{kphiD}).
Then we used the expansion of $I_0^2(t)$, in~(\ref{I02}),
and of~\cite[9.6.13]{AandS}
\begin{equation}
K_0(t)=-(\log(t/2)+\gamma)I_0(t)+\sum_{k=1}^\infty
\left(\frac{t^k}{2^k{}k!}\right)^2\sum_{n=1}^k\frac{1}{n}\,,
\label{K0}
\end{equation}
where $\gamma$ is Euler's constant, to obtain from~(\ref{Dseries})
and~(\ref{by6}) the closed form
\begin{equation}
h_k=\sum_{j=1}^k{k\choose j}^2{2j\choose j}\sum_{n=0}^{j-1}\frac{2}{k-n}
\label{his}
\end{equation}
which is a harmonic twist of the closed form~(\ref{A2893}) for the
hexagonal lattice integers $a_k$. Hence we obtain an
integer sequence for $H_k:=k!h_k/4$, with $k>0$, beginning with
\[1,\,13,\,263,\,7518,\,280074,\,12895572,\,707902740,\,45152821872,\,
3282497058384\ldots\]
Thanks to this sequence, we were able to evaluate all
the single integrals over $D(y)$ and $E(x)$ in this work
to 1200 decimal places, since we had very good control of
logarithmic singularities near the endpoints, as $y\to0$
and $x\to\infty$, respectively.

\subsection{A modular identity from quantum field theory}

A careful analysis~\cite{davdel} of the Dalitz plot
shows that the relation~(\ref{kphi}) between $k$ and $\phi$ implies
that~\cite[Eq.\ 5.17]{davdel}
\begin{equation}
Z(F(\sin\phi,k),k)=\frac13k^2\sin^3\phi
\label{plot}
\end{equation}
where $Z$ is \emph{Maple}'s {\tt JacobiZeta} function.
Combining our new finding~(\ref{E1E2}) with this, we obtain
from~\cite[17.4.38]{AandS} an elementary evaluation of
\begin{equation}
\frac{\sqrt3\pi}{4\,{\bf K}(k)}\sum_{n=0}^\infty\left(
\frac{4}{\left(q^{-3n-1}-q^{3n+1}\right)}-
\frac{4}{\left(q^{-3n-2}-q^{3n+2}\right)}\right)
=\frac12t^{-1}-\frac16t^{-3}
\label{qseries}
\end{equation}
where $q$ is the \emph{nome} associated with $k$
and $t=\tan(\phi/2)$
is determined algebraically by
\begin{eqnarray*}
r&=&\left(4k^2(1-k^2)\right)^{\frac13}\,,\quad
s=2k\sqrt{r+k^2}+2k^2+\frac{2(1-2k^2)r}{\sqrt{1+r+r^2}}-r\,,\\
t&=&\left(\frac{k-\sqrt{r+k^2}+\sqrt{s}}
{3k+\sqrt{r+k^2}-\sqrt{s}}\right)^{\frac12}\,.
\end{eqnarray*}
In particular, for $k=1/\sqrt2$ we obtain
$t=(2/\sqrt3-1)^{1/4}.$

A modular setting for this result is provided by Jacobi's identity
\begin{equation}
\sqrt{k}\,\theta_3(q)=\theta_2(q):=
\sum_{n=-\infty}^\infty q^{\left(n+\frac12\right)^2}
\label{theta2}
\end{equation}
with $\theta_3(q)$ related to ${\bf K}(k)$ by~(\ref{theta3}).
Summation of the Lambert series in~(\ref{qseries})
gives $\theta_2^3(q^{3/2})/\theta_2(q^{1/2})$, which is
a result known to Ramanujan. Thus we have proven
the \emph{modular identity}
\begin{equation}
{\sqrt3}\frac{\theta_2^3(q^{3/2})}
{\theta_3^2(q)\theta_2(q^{1/2})}
=t^{-1}-\frac13t^{-3}
\label{modular}
\end{equation}
where $c:=t^{-2}$ is the unique positive root of the polynomial
\begin{equation}
S(x)=3+8\left(1-2\frac{\theta_2^4(q)}{\theta_3^4(q)}\right)x+6x^2-x^4
\label{Joubert}
\end{equation}
and indeed $1\le c\le {3}$.
The real roots are $-\theta_3^2(q^{1/3})/\theta_3^2(q)$
and $3\,\theta_3^2(q^3)/\theta_3^2(q)$,
as has been known since Joubert and Cayley~\cite[(4.6.14)]{agm}.
Hence~(\ref{modular}) devolves to
\begin{equation}
\frac{\theta_2^3(q^{3/2})}
{\theta_2(q^{1/2})}
=\theta_3^2(q)\left\{\frac{\theta_3(q^3)}{\theta_3(q)}
-\frac{\theta_3^3(q^3)}{\theta_3^3(q)}\right\}.
\label{modular2}
\end{equation}

Such modular identities are machine-provable---in principle
and in practice---by computing that sufficiently
many terms of the Taylor series agree.
This is the so-called ``modular machine"~\cite[\S3]{garvan}.
In this case confirming 1000-term agreement is more than adequate
to prove~(\ref{modular2}); as takes seconds in \emph{Maple}.
A conventional proof can be pieced together from
\cite[Thm.\ 4.11]{agm}.
Our proof came from quite another source: the Lorentz
invariance of quantum field theory in two spacetime dimensions,
which enabled us to prove that~(\ref{E1}) and~(\ref{E2})
are identical.

\subsection{A discrete sum rule}

It seemed reasonable to try to prove the evaluation
for $t_{5,1}$ in one of its forms
\begin{eqnarray}
t_{5,1}=\int_{0}^{\frac13}\frac{D(y)\,{\rm d}y}{\sqrt{1-4y^2}}
&\stacktick{5}&\frac{1}{4}\,K_{15}K_{5/3}
\label{DJBconD}\\
t_{5,1}=\int_{0}^\infty\frac{E(x)\,{\rm d}x}{\sqrt{4+x^2}}
&\stacktick{5}&\frac{1}{4}\,K_{15}K_{5/3}
\label{DJBconE}
\end{eqnarray}
with $D(y)$ given by~(\ref{dee}) and $E(x)$ by~(\ref{eee}).

For example, by using~\cite[(2.35), p.\ 119]{expm2} one may deduce that
\begin{equation}
\int_0^{\pi/2}\int_0^{\pi/2}
\frac{{\rm d}\theta\,{\rm d}\phi}
{\sqrt{64-(16-\sin^2\phi)\sin^2\theta}}
=\frac{1}{8}\,K_{15}K_{5/3}
\label{hope}
\end{equation}
and then hope to relate some such recognizable
double integral to the single integrals in~(\ref{DJBconD})
and~(\ref{DJBconE}).

In fact, the integral forms~(\ref{DJBconD}) and~(\ref{DJBconE}) resisted
prolonged and intense efforts to find a proof. A break-through
came from the sum in~(\ref{t51sum}), which we rewrite as
\begin{equation}
\sum_{k=1}^\infty\frac{b_{k-1}}{64^k}
\stacktick{5}\frac{1}{16\pi^2}\,K_{15}K_{5/3}
\label{t51sumalt}
\end{equation}
whose right-hand side we shall relate to the
first of the generalized Watson
integrals~\cite{Joyce94,joyce-zucker}
\begin{equation}
W_j(z):=\frac{1}{\pi^3}\int_0^\pi\int_0^\pi\int_0^\pi
\frac{{\rm d}\theta_1\,{\rm d}\theta_2\,{\rm d}\theta_3}
{1-z\,w_j(\theta_1,\theta_2,\theta_3)}
\label{Watson}
\end{equation}
with
\begin{equation}
w_1(\theta_1,\theta_2,\theta_3)=\frac
{\cos\theta_1\cos\theta_2+\cos\theta_2\cos\theta_3+\cos\theta_3\cos\theta_1}{3}
\label{fcc}
\end{equation}
in the case of a \emph{face-centred cubic} (f.c.c.) lattice~\cite{Joyce71},
\begin{equation}
w_2(\theta_1,\theta_2,\theta_3)=\frac
{\cos\theta_1+\cos\theta_2+\cos\theta_3}{3}
\label{sc}
\end{equation}
in the case of a \emph{simple cubic} (s.c.) lattice~\cite{Joyce73},
\begin{equation}
w_3(\theta_1,\theta_2,\theta_3)=\cos\theta_1\cos\theta_2\cos\theta_3
\label{bcc}
\end{equation}
in the case of a \emph{body-centred cubic} (b.c.c.) lattice~\cite{Joyce71b} and
\begin{equation}
w_4(\theta_1,\theta_2,\theta_3)=
\frac{1
+\cos\theta_1\cos\theta_2+\cos\theta_2\cos\theta_3+\cos\theta_3\cos\theta_1}{4}
\label{diamond-green}
\end{equation}
in the case of a \emph{diamond} lattice~\cite{hijjawi}.

In 1971, Joyce gave the notable f.c.c.\
evaluation~\cite[Eq.\ 4]{Joyce71}
\begin{equation}
W_1(z)=\frac{12}{\pi^2}\,\frac{{\bf K}(k_+(z))\,{\bf K}(k_-(z))}{3+z}
\label{fcc71}
\end{equation}
with
\begin{equation}
k_\pm^2(z)=\frac12\pm\frac{2\sqrt3\,z}{(3+z)^{3/2}}
-\frac{\sqrt3}{2}\,\frac{(3-z)(1-z)^{1/2}}{(3+z)^{3/2}}\,.
\label{fcc71k}
\end{equation}
Our progress resulted from the intriguing observation that
at $z=\frac15$ this gives $k_-(1/5)=k_{15}$ and $k_+(1/5)=k_{5/3}$. Hence
we obtain
\begin{equation}
W_1\left(\frac15\right)=\frac{15}{4\pi^2}\,K_{15}K_{5/3}
\label{obs5}
\end{equation}
which reduces~(\ref{t51sumalt}) to the discrete sum rule
\begin{equation}
\sum_{k=1}^\infty\frac{b_{k-1}}{64^k}
\stacktick{5}
\sum_{k=1}^\infty\frac{f_{k-1}}{60^k}
\label{conj5w}
\end{equation}
where the f.c.c.\ lattice integers
\begin{equation}
f_k:=\frac{4^k}{\pi^3}\int_0^\pi\int_0^\pi\int_0^\pi
\left(\cos\theta_1\cos\theta_2+\cos\theta_2\cos\theta_3
+\cos\theta_3\cos\theta_1\right)^k\,{\rm d}\theta_1\,{\rm d}\theta_2\,
{\rm d}\theta_3
\label{A2899}
\end{equation}
give the Taylor coefficients of the expansion
\begin{equation}
W_1(z)=\sum_{k=0}^\infty f_k\left(\frac{z}{12}\right)^k
\label{fcc-exp}
\end{equation}
for $|z|\le1$.
In the next subsection,
we show that~(\ref{conj5w}) derives from a more
general set of sum rules. Then, in subsection~\ref{cubic-mod},
we shall expose a cubic modular equation, implicit in~(\ref{fcc71k}).

\subsection{Sum rules for diamond and cubic lattice integers}
\label{proof}

The first few terms in the
sequence for the f.c.c.\ lattice integers~(\ref{A2899}) are
\begin{equation}
1,\,0,\,12,\,48,\,540,\,4320,\,42240,\,403200,\,4038300,\,40958400,\,
423550512,\,4434978240
\label{A2899s}
\end{equation}
for $n=0\ldots11$. The values up to $f_8=4038300$ are recorded
in~\cite[Table 1]{Lcluster} and the next integer, $f_9=40958400$,
was given by Domb~\cite{crystals} and
recorded\footnote{See {\tt
http://www.research.att.com/\~{ }njas/sequences/A002899}~.}
in entry A2899 of the on-line version of~\cite{EIS},
which provided us with no closed formula for $f_k$.
By way of contrast, the s.c., b.c.c.\ and diamond lattice
expansions
\begin{eqnarray}
W_2(z)&=&\sum_{k=0}^\infty{2k\choose k}a_k\left(\frac{z}{6}\right)^{2n}
\label{sc-exp}\\
W_3(z)&=&\sum_{k=0}^\infty{2k\choose k}^3\left(\frac{z}{8}\right)^{2n}
\label{bcc-exp}\\
W_4(z)&=&\sum_{k=0}^\infty b_k\left(\frac{z}{16}\right)^n
\label{diamond-exp}
\end{eqnarray}
lead to explicit expressions for the integer sequences
in entries A2896, A2897 and A2895,
respectively, of the on-line version of~\cite{EIS}.
As in the f.c.c.\ case~(\ref{fcc-exp}), expansions~(\ref{sc-exp})
to~(\ref{diamond-exp}) are valid for $|z|\le1$.
For convenience, we repeat here the closed forms
\begin{equation}
a_k=\sum_{j=0}^k{k\choose j}^2{2j\choose j}\,,\quad\quad
b_k=\sum_{j=0}^k{k\choose j}^2{2k-2j\choose k-j}{2j\choose j}\,,
\label{A2893-A2895}
\end{equation}
for the hexagonal and diamond lattice integers, previously
given in~(\ref{A2893}) and~(\ref{A2895}). The hexagonal lattice
integers $a_k$ appear in the s.c.\ lattice integers
${2k\choose k}a_k$ of expansion~(\ref{sc-exp}). We also note
the terminating hypergeometric series
\[b_k={2k\choose k}\;{}_4{\rm F}_3
\left(\begin{array}{ccccc}\frac12,-k,-k,-k\\
1,1,-k+\frac12\end{array}\bigg|1\right)\]
for the diamond lattice integers. Likewise
\[a_k={}_3{\rm F}_2
\left(\begin{array}{cccc}\frac12,-k,-k\\
1,1\end{array}\bigg|4\right).\]

We were able to derive the closed form
\begin{equation}
f_k=\sum_{j=0}^k{k\choose j}(-4)^{k-j}b_j
\label{fcc-closed}
\end{equation}
for the f.c.c.\ lattice integers,
by noting the similarity between~(\ref{fcc}) and~(\ref{diamond-green}),
which leads to the functional relationship
\begin{equation}
\frac{1}{4N+4}\,W_4\left(\frac{4}{N+1}\right)=
\frac{1}{4N}\,W_1\left(\frac{3}{N}\right)
\label{W4-W1}
\end{equation}
between the Green functions for the diamond and f.c.c.\ lattices.
Then for $N\ge3$ we may expand each side, to obtain the sum rule
\begin{equation}
\sum_{k=1}^\infty\frac{b_{k-1}}{(4N+4)^k}=
\sum_{k=1}^\infty\frac{f_{k-1}}{(4N)^k}
\label{only-connect}
\end{equation}
and derive the
closed form~(\ref{fcc-closed}) for the f.c.c.\ lattice integers
defined in~(\ref{A2899}) by further expanding the left-hand side in
powers of $1/N$.

Setting $N=3$ in~(\ref{only-connect}), we prove the sum rule
\begin{equation}
\sum_{k=1}^\infty\frac{b_{k-1}}{16^k}=
\sum_{k=1}^\infty\frac{f_{k-1}}{12^k}=
\frac{1}{12}W_1(1)=\frac{1}{4\pi^2}\,K_{3}K_{1/3}
\label{conj8w}
\end{equation}
from Watson's evaluation~\cite{watson1} of $W_1(1)$, at the
singular value $k_3$.

Setting $N=15$ in~(\ref{only-connect}), we prove the sum rule
\begin{equation}
\sum_{k=1}^\infty\frac{b_{k-1}}{64^k}=
\sum_{k=1}^\infty\frac{f_{k-1}}{60^k}=
\frac{1}{60}W_1\left(\frac15\right)=\frac{1}{16\pi^2}\,K_{15}K_{5/3}
\label{conj5w-bis}
\end{equation}
at the singular value $k_{15}$. Hence we have proven~(\ref{conj5w})
and all the other forms of our initial conjecture~(\ref{t51})
for the moment $t_{5,1}$. By taking up to 4 differentials of
$W_1(3/N)$, before setting $N=15$,
one may also prove the evaluations of $t_{5,3}$ in~(\ref{t53})
and $t_{5,5}$ in~(\ref{t55}),
using the evaluations of the elliptic integrals ${\bf E}(k_{15})$
and ${\bf E}(k_{5/3})$ given explicitly in the talk of footnote~3 and
derivable from identities in~\cite{agm}.

These eventual proofs of our conjectures for the moments $t_{5,2k+1}$
came from noting a parenthetical remark
in~\cite[p.\ 601]{Joyce73} to the effect that the Green function
for the diamond lattice is given by a product of complete
elliptic integrals. It will be useful to consider the generating function
$\widetilde{D}$ which has the explicit form
\begin{equation}
\widetilde{D}(y):=\frac{2\,{\bf K}
\left(\sqrt{\frac{16y^3}{(1+3y)(1-y)^3}}\right)}
{\pi\sqrt{(1+3y)(1-y)^3}}
=\frac{1}{{\rm AGM}(\sqrt{(1-3y)(1+y)^3},\sqrt{(1+3y)(1-y)^3})}
\label{dee-tilde}
\end{equation}
with a complete elliptic integral that is complementary
to that in~(\ref{dee}) for $D(y)$. For $|y|<\frac13$,
we have the expansion
\begin{equation}
\widetilde{D}(y)
={\rm HeunG}\left(9,3;1,1,1,1;9y^2\right)
=\sum_{k=0}^\infty a_k y^{2k}
\label{dee-tilde-hex}
\end{equation}
in terms of the hexagonal lattice integers in~(\ref{A2893-A2895}).

The full story, for the f.c.c., s.c.,
b.c.c. and diamond lattices, is then provided by the four identities
\begin{eqnarray}
\widetilde{D}^2\left(\sqrt{-x}\right)=
\left(\sum_{k=0}^\infty \,a_k\,(-x)^k\right)^2
&=&\sum_{k=0}^\infty \,f_k\,\frac{ x^k}{(1+3x)^{2k+2}}
\label{fcc-dee}\\
&=&\sum_{k=0}^\infty\,a_k\,\frac{{2k\choose k}\left(-x(1+x)(1+9x)\right)^k}
{((1-3x)(1+3x))^{2k+1}}
\label{sc-dee}\\
&=&\sum_{k=0}^\infty\frac{\left({2k\choose k}x^k\right)^3}
{((1+x)^3(1+9x))^{k+\frac12}}
\label{bcc-dee}\\
&=&\sum_{k=0}^\infty\,b_k\,\frac{ x^k}{((1+x)(1+9x))^{k+1}}
\label{diamond-dee}
\end{eqnarray}
which are valid for sufficiently small $x$ and were obtained
by simplification of formulae in~\cite{Joyce71,Joyce71b,Joyce73}.

At bottom, all 4 results~(\ref{fcc-dee}) to~(\ref{diamond-dee})
originate from the first paper~\cite{bailey1936a}
of Wilfrid Norman Bailey's adroit series on
infinite integrals involving Bessel
functions, in much the same way that our proof of~(\ref{djbwnb})
resulted from his second paper~\cite{bailey1936}.
In~\cite[Eq.\ 8.1]{bailey1936a}, Bailey showed that
$\int_0^\infty J_\mu(a t)J_\nu(b t)J_\rho(c t)\,{\rm d}t$
is given by a product of ${}_2{\rm F}_1$ hypergeometric functions.
In~\cite{Joyce71,Joyce71b,Joyce73},
Joyce used this result
to obtain diamond and cubic lattice Green functions, in three
spatial dimensions, from the \emph{square} of the Green
function~(\ref{dee-tilde}) for the hexagonal (or ``honeycomb")
lattice in two spatial dimensions, for which we have given a {\tt HeunG} form
in~(\ref{dee-tilde-hex}) equivalent to that
given in~\cite[Eq.\ 4]{guttmann}.
With $x\ge0$, the explicit form of $E$ in~(\ref{eee})
provides an analytic continuation for
\begin{equation}
\widetilde{D}\left(\sqrt{-x}\right)
=\frac{3\sqrt3}{\pi}\,E\left(3\sqrt{x}\right)\,.
\label{eee-dee}
\end{equation}

The singular value $k_{15}$, obtained from the
diamond and f.c.c.\ lattice sums in~(\ref{conj5w-bis}),
also appears in the s.c.\ and b.c.c.\ lattice sums
\begin{eqnarray}
\sum_{k=0}^\infty\frac{{2k\choose k}}{(-45)^k}\,a_k
&=&\frac{3\sqrt5}{2\pi^2}\,K_{15}K_{5/3}
\label{sc-sum}\\
\sum_{k=0}^\infty\frac{{2k\choose k}^3}{2^{4k}(\sqrt5+1)^{8k}}
&=&\frac{4}{\pi^2}\,K_{15}^2
\label{bcc-sum}
\end{eqnarray}
obtained by setting $x=x_{15}:=3-\frac43\sqrt5\approx0.018576$
in~(\ref{sc-dee}) and~(\ref{bcc-dee}).
Setting $x=x_{15}$ in~(\ref{eee-dee}),
we obtain the singular value $k_{5/3}=\sin(\alpha(3\sqrt{x_{15}}))$ from
definition of the angle $\alpha(x)$ in~(\ref{alpha-x}).
This is confirmed by the formula for $\tan(2\alpha(x))$ in~(\ref{tan2}).
We remark that~(\ref{sc-sum}) follows from
the functional relationship
given in~\cite[Eq.\ 3.23]{joyce-zucker}
and that~(\ref{bcc-sum}) follows from~\cite[Table 5.2a ($N=15$)]{agm}
and the Clausen product in~\cite[Th.\ 5.7(a)(i), p.\ 180]{agm}.
We now show how to obtain other singular values from lattice Green
functions, using a modular identity.

\subsection{Cubic modular equations}
\label{cubic-mod}

The \emph{cubic modular equation}~\cite[Th.\ 4.1, p.\ 110]{agm}
\begin{equation}
\theta_4(q)\theta_4(q^3)+\theta_2(q)\theta_2(q^3)
=\theta_3(q)\theta_3(q^3)
\label{cubic}
\end{equation}
relates instances of the Jacobi functions $\theta_2$, in~(\ref{theta2}),
and $\theta_3$, in~(\ref{theta3}), with \emph{nomes} $q$ and $q^3$,
to corresponding instances of
\begin{equation}
\sqrt{k^\prime}\,\theta_3(q)=\theta_4(q):=
\sum_{n=-\infty}^\infty(-1)^n q^{n^2}
\label{theta4}
\end{equation}
with $k^\prime:=\sqrt{1-k^2}$ and \emph{nome} $q$ associated with $k$.
If we associate $q^3$ with the complete elliptic argument $l$,
then~(\ref{cubic}) gives the identity~\cite[Eqn.\ 4.2.6]{agm}
$\sqrt{k^\prime l^\prime}+\sqrt{k\,l}=1$. The results of the
previous subsection follow from the notable circumstance that~(\ref{fcc71k})
gives
\begin{equation}
\sqrt{k_+^\prime(z)k_-^\prime(z)}+\sqrt{k_+(z)k_-(z)}-1=0
\label{cubic-z}
\end{equation}
which may be proven symbolically, by setting $z=1-t^2$ and
denoting the left-hand side of~(\ref{cubic-z}) by $y(t)$.
Then $y(t)$ is analytic on the closed unit disk, $|t|\le1$, and
\emph{Maple} computes an algebraic equation of the form
$y(t)\,P(y(t),t)=0$ with $P(0,0)$ non-zero. This proves that $y(t)$
vanishes in some neighbourhood of $t=0$ and hence for $|t|\le1$.

The resulting modular identities for the Green functions of cubic
lattices are most conveniently obtained from the Green function
$W_4$, for the diamond lattice, with a parametric solution
\begin{equation}
W_4(z_4)=\theta_3^2(q)\theta_3^2(q^3)\,,\quad\quad
z_4:=\frac{1-4\eta^2}{1-\eta^2}\,,\quad\quad
\eta:=\frac{3\theta_3^4(q^3)-\theta_3^4(q)}{3\theta_3^4(q^3)+\theta_3^4(q)}\,,
\label{diamond-param}
\end{equation}
corresponding to the series solution~(\ref{diamond-dee}) at
$x=(1-2\eta)/(3+6\eta)$. Then the Green functions for the
f.c.c., s.c.\ and b.c.c. lattices
are given by the functional relationships
\begin{equation}
\frac43(1-\eta^2)\,W_1(z_1)=
\frac2{3\eta}(1-\eta^2)\,W_2(z_2)=
\sqrt{\frac{3-3\eta}{1+\eta}}\,W_3(z_3)=W_4(z_4)
\label{cubic-w123}
\end{equation}
with the arguments
\begin{equation}
z_1=1-4\eta^2\,,\quad\quad
z_2=\sqrt{z_1\left(1-\frac{1}{\eta^2}\right)}\,,\quad\quad
z_3=\frac{1-2\eta}{2+2\eta}\,\sqrt{z_4}\,,
\label{cubic-z123}
\end{equation}
obtained from~(\ref{fcc-dee}) to~(\ref{bcc-dee}), respectively,
by the substitution $x=(1-2\eta)/(3+6\eta)$ that
gave~(\ref{diamond-param}).
We note the alternative b.c.c.\
parameterization~\cite[Th.\ 5.7(a)(i), p.\ 180]{agm},
\begin{equation}
W_3(z_3)={}_3{\rm F}_2
\left(\begin{array}{cccc}\frac12,\frac12,\frac12\\
1,1\end{array}\bigg|z_3^2\right)
=\theta_3^4(q^3)\,,\quad\quad
z_3=2l\sqrt{1-l^2}\,,\quad\quad
l:=\frac{\theta_2^2(q^3)}{\theta_3^2(q^3)}\,.
\label{bcc-param}
\end{equation}
The equivalence of the forms
for $z_3$ in~(\ref{cubic-z123}) and~(\ref{bcc-param}) results
from the modular identity
\begin{equation}
z_3^2=\frac{(1-2\eta)^2}{(2+2\eta)^2}\,
\frac{1-4\eta^2}{1-\eta^2}=4l^2(1-l^2)=(2ll^\prime)^2
\label{cubic-bis}
\end{equation}
with $\eta$ defined in~(\ref{diamond-param}) and $l$ in~(\ref{bcc-param}).
To prove~(\ref{cubic-bis}), we combined the cubic modular
identity~(\ref{cubic}) with the Joubert--Cayley
result~\cite[(4.6.14)]{agm} $S(3\theta_3^2(q^3)/\theta_3^2(q))=0$,
where $S$ is defined in~(\ref{Joubert}). Alternatively, we may again apply
the ``modular machine"~\cite[\S3]{garvan}.\footnote{Striking cubic modular
equations ($q \mapsto q^3$ or $N \mapsto 9N$), originating with Ramanujan,
are explored in~\cite[\S4.7]{agm}. In particular there are attractive cubic
recursions for the \emph{cubic multiplier} $M=\sqrt{(1+\eta)/(3-3\eta)}
=\theta_3^2(q^3)/\theta_3^2(q)$, as occurs in~(\ref{cubic-w123}).}

Noting that the lattices sums
in~(\ref{conj8w}) and~(\ref{conj5w-bis})
yield the singular values $k_3$ and $k_{15}$,
from rational summands, we wondered
if any other singular value might be obtained from
a lattice Green function in such a neat manner. We know
from Watson's classic work~\cite{watson1}
that $W_3(1)$ yields the singular value
$k_1=1/\sqrt2$ while $W_2(1)$ yields the singular value
$k_6=(\sqrt3-\sqrt2)(2-\sqrt3)$, as noted in~\cite{expm2}.
Moreover,~(\ref{bcc-param}) is equivalent to
$W_3(2k_{N}k_{N}^\prime)=4K_N^2/\pi^2$, for $q^3=\exp(-\pi\sqrt{N})$.
Hence $W_3(z_3)$ yields the singular values $k_1,\,k_3,\,k_7$
for the rational arguments $z_3=1,\,\frac12,\,\frac18$, respectively.

Prompted by the sum over s.c.\ lattice integers in~(\ref{sc-sum}), we sought
further examples, in which a sum over rational numbers might lead to
a singular value $k_N$. We found 5 new results for $W_2$,
which appear to exhaust the cases with rational summands.
These occur with $N/3=7,11,19,31,59$, for which
we obtained
\begin{eqnarray}
\sum_{k=0}^\infty\frac{{2k\choose k}}{(-108)^k}\,a_k
&=&\frac{6}{\pi^2}\left(3\sqrt3-\sqrt{21}\right)\,K_{21}K_{7/3}
=\frac37\,G(21)
\label{K21}\\
\sum_{k=0}^\infty\frac{{2k\choose k}}{(-396)^k}\,a_k
&=&\frac{6}{\pi^2}\left(3\sqrt{33}-5\sqrt{11}\right)\,K_{33}K_{11/3}
=\frac{\sqrt3}{\sqrt{11}}\,G(33)
\label{K33}\\
\sum_{k=0}^\infty\frac{{2k\choose k}}{(-2700)^k}\,a_k
&=&\frac{30}{\pi^2}\left(3\sqrt{57}-13\sqrt3\right)\,K_{57}K_{19/3}
=\frac{15}{19}\,G(57)
\label{K57}\\
\sum_{k=0}^\infty\frac{{2k\choose k}}{(-24300)^k}\,a_k
&=&\frac{90}{\pi^2}\left(39\sqrt3-7\sqrt{93}\right)\,K_{93}K_{31/3}
=\frac{45}{31}\,G(93)
\label{K93}\\
\sum_{k=0}^\infty\frac{{2k\choose k}}{(-1123596)^k}\,a_k
&=&\frac{69}{8\pi^2}\left(\sqrt3-1\right)^9\sqrt{59}\,K_{177}K_{59/3}
=\frac{23\sqrt3}{\sqrt{59}}\,G(177)
\label{K177}
\end{eqnarray}
with reductions to $\Gamma$ values given by
\begin{equation}
G(N)=\frac{1}{2\sqrt2\pi}\prod_{n=1}^{4N-1}
\left[\Gamma\left(\frac{n}{4N}\right)\right]^{\frac14\,(-4N|n)}
\label{Gamma}
\end{equation}
where $\Gamma(n/(4N))$ contributes to the product if $n$ is coprime to $4N$
and then occurs with an exponent $\pm\frac14$, according as the sign of the
Legendre--Jacobi--Kronecker symbol $(-4N|n)$. We remark that
$p=7,11,19,31,59$ are the only primes for which $N=3p$ is a
\emph{disjoint discriminant} of \emph{type one}, as considered
in~\cite[Eq.\ 9.2.8, p.\ 293]{agm}.

We may obtain other singular values
by choosing the argument of $W_4$ in~(\ref{diamond-param}) to be an
appropriate \emph{algebraic} number. For example, the sums
\begin{eqnarray}
\sum_{k=0}^\infty\frac{b_k}
{(12+4\sqrt{13})^{2k}}
&=&\frac{4}{\pi^2}\,K_{39}K_{13/3}
\label{K39}\\
\sum_{k=0}^\infty\frac{b_k}
{(3\sqrt3+5)^{2k}(6\sqrt3+4\sqrt7)^{2k}}
&=&\frac{4}{\pi^2}\,K_{105}K_{35/3}
\label{K105}
\end{eqnarray}
over the diamond lattice integers $b_k$ in~(\ref{A2893-A2895})
have relatively simple surds in their summands, obtained
by setting $q=\exp(-\pi\sqrt{13/3})$ and $q=\exp(-\pi\sqrt{35/3})$
in the parametric solution~(\ref{diamond-param}).
We may also obtain quartic values of $z_4$ and evaluations
like~(\ref{K105}) at the even singular values $k_N$
with $N/6=3,5,7,13,17$. We remark that $p=5,7,13,17$ are
the only primes for which $N/2=3p$ is a
\emph{disjoint discriminant} of \emph{type two}, as considered
in~\cite[Eq.\ 9.2.9, p.\ 293]{agm}.

\subsection{Integral sum rules}

Unfortunately, the discrete sum rule~(\ref{conj8w}), at the singular
value $k_3$, does not prove
conjecture~(\ref{DJB3a}) but instead converts it to
the conjectural integral sum rule
\begin{equation}
\int_0^\infty t\,I_0(t)\left(I_0(t)K_0(2t)-\frac13\,I_0(2t)K_0(t)\right)
K_0^2(t)\,{\rm d}t
\stackconj{8}0
\label{connect8}
\end{equation}
where the term containing $I_0(2t)$ is now proven to
yield the singular value $k_3$, whose appearance in
the term containing $K_0(2t)$ was
conjectured in~(\ref{DJB3a}).

It looks to be an even tougher proposition
to prove the sum rule
\begin{equation}
\int_0^\infty
t\,I_0(t)\left(K_0(t)-\frac{2\pi}{\sqrt{15}}I_0(t)\right)K_0^3(t)\,{\rm d}t
\stackconj{2}0
\label{sum}
\end{equation}
for which we have now obtained two representations
in terms of integrals of complete elliptic integrals, namely
\begin{eqnarray}
\int_{0}^{\frac13}\frac{D(y)}{\sqrt{1-4y^2}}
\left(\,{\rm arctanh}\left(\sqrt{\frac{1-2y}{1+2y}}\right)
-\frac{\pi}{\sqrt{15}}\right)\,{\rm d}y
&\stackconj{2}&0
\label{sumD}\\
\int_{0}^\infty\frac{E(x)}{\sqrt{4+x^2}}
\left(\,{\rm arcsinh}\left(\frac{x}{2}\right)-\frac{\pi}{\sqrt{15}}
\right)\,{\rm d}x
&\stackconj{2}&0
\label{sumE}
\end{eqnarray}
with $D(y)$ given in~(\ref{dee}), $E(x)$ given in~(\ref{eee}) and
the 3-loop sunrise diagram $s_{5,1}$ appearing
via the more demanding logarithmic terms
in~(\ref{Ds51}) and~(\ref{Es51}).

As a companion to the proven evaluation~(\ref{jonc30})
and the discrete sum rule~(\ref{DJBdisE3}), we present
\begin{equation}
\int_0^{\pi/2}\frac{{\bf K}(\sin\theta)}
{\sqrt{1+3\sin^2\theta}}
\left({\rm arcsinh}(2\tan\theta)-\frac{\pi}{\sqrt3}\right)
\,{\rm d}\theta\
\stackconj{10}0
\label{DJBrelE3}
\end{equation}
as our penultimate conjecture, also checked to 1200 decimal places.
If the mathematics of sum rule~(\ref{DJBrelE3})
might be elucidated at the singular value $k_3$,
then there might be some hope for a proof of the quantum field theory
result~(\ref{sum}), at the singular value $k_{15}$.

\subsection{The even moment $c_{5,0}$}

We were unable to derive single integrals of elliptic integrals
for the even moments $c_{5,2k}$. A double-integral representation
is readily available by setting $w=0$ in the \emph{Aufbau}
\begin{eqnarray}
\overline{S}_6(a,b,c,d,e,w)
&=&\int_{-\infty}^\infty
\overline{S}_4(a,b,c,v)\overline{S}_3(d,e,v+w)\,{\rm d}v\nonumber\\
&=&\int_{-\infty}^\infty\int_{-\infty}^\infty
\overline{S}_3(a,b,u)\overline{S}_2(c,u+v)\overline{S}_3(d,e,v+w)
\,{\rm d}u\,{\rm d}v
\label{noess6}
\end{eqnarray}
to obtain
\begin{equation}
\overline{V}_5(a,b,c,d,e)=
\frac{\pi}{2}\int_{-\infty}^\infty\int_{-\infty}^\infty
\frac{\overline{S}_3(a,b,u)\overline{S}_3(d,e,v)}{\sqrt{c^2+(u+v)^2}}
\,{\rm d}u\,{\rm d}v
\label{novee5}
\end{equation}
with $\overline{S}_3$ given by~(\ref{noess3}).
Setting the 5 parameters to unity and making the transformations
$u=2\tan\theta$ and $v=2\tan\phi$, we obtain
\begin{equation}
c_{5,0}=\frac{\pi}{2}
\int_{-\pi/2}^{\pi/2}
\int_{-\pi/2}^{\pi/2}
\frac{{\bf K}(\sin\theta)\,{\bf K}(\sin\phi)}
{\sqrt{\cos^2\theta\cos^2\phi+4\sin^2(\theta+\phi)}}
\,{\rm d}\theta\,{\rm d}\phi\,.
\label{cee50}
\end{equation}
The higher even moments are obtained by suitable differentiations
of~(\ref{novee5}) with respect to $c$, before setting $c=1$.

\section{Six Bessel functions}

We are now equipped to write the odd moments
$t_{6,1}$, $s_{6,1}$ and $c_{6,1}$ as
single or double integrals over complete elliptic
integrals, with integrands that are computable
with great efficiency, using the exponentially
fast process of the arithmetic-geometric mean,
discovered by Lagrange, around 1784, and independently
by Gauss, at the age of 14, in 1791~\cite{agmhist,agm}.

\subsection{The odd moments $t_{6,2k+1}$}

We begin by folding $D_4$, in~(\ref{dee4}),
with $T_4$, in~(\ref{tee4}), to determine
\begin{eqnarray*}
T_6(u,a,b,c,d,v)&:=&
\int_0^\infty t\,J_0(u t)K_0(a t)K_0(b t)K_0(c t)K_0(d t)J_0(v t)\,{\rm d}t\\
&=&\int_{a+b+c}^\infty 2w\,D_4(a,b,c,w)T_4(u,d,w,v)\,{\rm d}w
\end{eqnarray*}
where we group the 3 internal lines with masses $a,b$ and $c$ to have
a total momentum with norm $w^2$.
Setting $a=b=c=d=1$,
$u=v={\rm i}$ and $w=1/y$, we obtain one instance of $D(y)$, from
its definition $D(y):=2D_4(1,1,1,1/y)/y$, and another, somewhat
surprisingly, from the novel result $D(y)=6T_4({\rm i},1,1/y,{\rm i})/y$
in~(\ref{by6}). Hence we obtain
\begin{equation}
t_{6,1}:=\int_0^\infty t\,I_0^2(t)K_0^4(t)\,{\rm d}t
=\frac13\int_{0}^\frac13\frac{D^2(y)}{2y}\,{\rm d}y
\label{t61}
\end{equation}
with an initial factor of $\frac13$ arising from~(\ref{E1E2bis}),
in the special case~(\ref{kphi}).
This is a rather efficient representation of $t_{6,1}$,
which delivers 1200 decimal places in two minutes, using \emph{Pari-GP}.
Similarly, we may derive a one-dimensional integral for
higher odd moments by differentiations of $D_4(a,b,c,w)$
with respect to one of its masses, before going to the equal-mass point.
More conveniently, one may use the summation
\[t_{6,2k+1}=\sum_{n=0}^\infty{2n\choose n}
\frac{c_{4,2k+2n+1}}{(2^n{}n!)^2}\]
that follows from~(\ref{I02}). By this means, we found that
$\frac{85}{72}t_{6,3}-\frac{1}{36}t_{6,1}$ reproduces the value
of $t_{6,5}$ to 1200 decimal places.

\subsection{The odd moments $s_{6,2k+1}$}

We were unable to derive a one-dimensional integral
over {\tt agm} functions for $s_{6,1}$, though we shall
conjecture such an integral, in the next subsection.
Here, the best that
we can do comes from using the folding
\[S_6(a,b,c,d,e,w)
=\int_{a+b+c}^\infty2u\,D_4(a,b,c,u)S_4(d,e,u,w)\,{\rm d}u\]
which leads to a choice of integrals for
\[S_4(d,e,u,w)
=\int_{d+e+u}^\infty2v\,D_4(d,e,u,v)S_2(v,w)\,{\rm d}v\\
=\int_{d+e}^\infty2v\,D_3(d,e,v)S_3(v,u,w)\,{\rm d}v\]
with the first form involving an {\tt agm} and the second an {\tt arctanh}
procedure. The latter is more convenient, since it yields
the rectangular double integral
\begin{equation}
s_{6,1}=\int_0^\frac13D(y)\int_0^\frac12
\frac{4z\,{\rm arctanh}\sqrt{A_-/A_+}}{\sqrt{(1-4z^2)A_+A_-}}
\,{\rm d}z\,{\rm d}y
\label{s61}
\end{equation}
with $A_\pm:=(y\pm z)^2-y^2z^2$, obtained by transforming to
$y=1/u$ and $z=1/v$, and $D(y)$ given in~(\ref{dee}).
We remark that $A_+$ is positive, within the rectangle
of integration, and that when $A_-$ is negative an analytic continuation
of {\tt arctanh} to {\tt arctan} keeps the integrand real and positive.
Higher odd moments may be obtained by differentiations with respect to $w$,
before setting $w^2=-1$. More conveniently, we may use the summation
\begin{equation}
s_{6,2k+1}=\sum_{n=0}^\infty\frac{c_{5,2k+2n+1}}{(2^n{}n!)^2}
\label{s61sum}
\end{equation}
since $c_{5,1}$ and $c_{5,3}$ were computed by Broadhurst
to 200,000 decimal places (see footnote 3) and higher moments
are (conjecturally) determined by them, using~(\ref{c55})
and the appropriate recursion from~(\ref{littlecrec}).
By this means, we found that
$\frac{85}{72}s_{6,3}-\frac{1}{36}s_{6,1}$ reproduces the value
of $s_{6,5}$ to 1200 decimal places.

\subsection{Sum rules}

Our final conjecture is that there is a infinite tower of sum rules
relating moments in which powers of $K_0$ are replaced
by corresponding powers of $\pi I_0$. We were alerted to this possibility
by the sum rule
\begin{equation}
\int_0^\infty \left(\pi^2I_0^2(t)-K_0^2(t)\right)K_0^2(t)\,{\rm d}t=0
\label{t40sum}
\end{equation}
proven in~(\ref{t40}).
For each pair of integers $(n,k)$ with $n\ge2k\ge2$ we conjecture that
\begin{equation}
Z_{2n,n-2k}:=\sum_{m=0}^{\lfloor n/2\rfloor}(-1)^m{n\choose 2m}
\int_0^\infty t^{n-2k}[\pi I_0(t)]^{n-2m}[K_0(t)]^{n+2m}\,{\rm d}t
\stackconj{11}0
\label{conj11}
\end{equation}
with the vanishing of $Z_{4,0}$ proven in~(\ref{t40sum}), in the case
with $n=2$ and $k=1$.

With 6 Bessel functions in play, the sum rule
\begin{equation}
Z_{6,1}:=\pi\int_0^\infty t\,I_0(t)\left(\pi^2I_0^2(t)-3K_0^2(t)\right)
K_0^3(t)\,{\rm d}t\stackconj{11}0\
\label{u61}
\end{equation}
relates a pair of odd moments, giving the conjectural evaluation
\begin{equation}
s_{6,1}\stackconj{11}\frac{\pi^2}{3}
\int_0^\infty t\,I_0^3(t)K_0^3(t)\,{\rm d}t\,.
\label{u61alt}
\end{equation}
At first sight, this seems to be hard to check,
at high precision, because it involves the slowly convergent moment
\[\int_0^\infty t\,I_0^3(t)K_0^3(t)\,{\rm d}t=\sum_{k=0}^\infty
a_k\frac{c_{3,2k+1}}{(2^k{}k!)^2}\]
with an integrand of order $1/t^2$, at large $t$,
and a summand of order $1/k^2$, at large $k$, coming
from the hexagonal lattice integers $a_k$ in~(\ref{A2893}).
However, we were able to exploit the
integral representation
\[\frac{c_{3,2k+1}}{(2^k{}k!)^2}=\int_0^\frac13D(y)y^{2k}\,{\rm d}y\]
to obtain
\begin{equation}
\int_0^\infty t\,I_0^3(t)K_0^3(t)\,{\rm d}t
=\int_0^\frac13D(y)\widetilde{D}(y)\,{\rm d}y
\label{u61int}
\end{equation}
as an integral over a pair of {\tt agm} functions, with $D(y)$
given by~(\ref{dee}) and $\widetilde{D}(y)$ by~(\ref{dee-tilde}).
Then conjecture~(\ref{u61}) is equivalent to the evaluation
\begin{equation}
s_{6,1}:=\int_0^\infty t\,I_0(t)K_0^5(t)\,{\rm d}t
\stackconj{11}
\frac{\pi^2}{3}\int_0^\frac13D(y)\widetilde{D}(y)\,{\rm d}y
\label{s61conj}
\end{equation}
which was confirmed to 1200 decimal places,
by setting $k=0$ in~(\ref{s61sum}) to compute $s_{6,1}$
and by using \emph{Pari-GP} to evaluate the integral
over $y$. We find it remarkable that the complicated
double integral 4-loop sunrise diagram in~(\ref{s61}) seems to be reducible
to the attractive single integral in~(\ref{s61conj}), by removing a
factor $\frac13\pi^2$.

With 8 Bessel functions in play, conjecture~(\ref{conj11}) gives a pair
of sum rules. {}From the vanishing of $Z_{8,0}$ and $Z_{8,2}$, we obtain
\begin{eqnarray}
c_{8,0}&
\stackconj{11}&
\pi^2\int_0^\infty I_0^2(t)
\left(6K_0^2(t)-\pi^2I_0^2(t)\right)K_0^4(t)\,{\rm d}t
\label{c80}\\
c_{8,2}
&\stackconj{11}&
\pi^2\int_0^\infty t^2 I_0^2(t)
\left(6K_0^2(t)-\pi^2I_0^2(t)\right)K_0^4(t)\,{\rm d}t
\label{c82}
\end{eqnarray}
again noting the slow convergence of these integrals and of the
equivalent sums
\begin{eqnarray}
c_{8,0}
&\stackconj{11}&
\sum_{k=0}^\infty\left(\frac{\pi}{2^k{}k!}\right)^2\left(
{2k\choose k}6c_{6,2k}-\pi^2b_k c_{4,2k}\right)
\label{c80sum}\\
c_{8,2}
&\stackconj{11}&
\sum_{k=0}^\infty\left(\frac{\pi}{2^k{}k!}\right)^2\left(
{2k\choose k}6c_{6,2k+2}-\pi^2b_k c_{4,2k+2}\right)
\label{c82sum}
\end{eqnarray}
which involve the diamond lattice integers $b_k$ in~(\ref{A2895})
or~(\ref{A2893-A2895}).

Conjecture~(\ref{conj11}) gives novel evaluations
of $c_{4n,2k}$ and $s_{4n+2,2k+1}$, for $n>k\ge0$.
We remark, however, that~(\ref{c80sum}) and~(\ref{c82sum})
do not exhaust the integer relations for moments with 8 Bessel functions.
We also found that the ratio
\[\frac
{\int_0^\infty t\,I_0^2(t)K_0^6(t)\,{\rm d}t}
{\int_0^\infty t\,I_0^4(t)K_0^4(t)\,{\rm d}t}
=\frac
{\sum_{k=0}^\infty{2k\choose k}\frac{c_{6,2k+1}}{(2^k{}k!)^2}}
{\sum_{k=0}^\infty b_k\frac{c_{4,2k+1}}{(2^k{}k!)^2}}\]
coincides with $\frac{9}{14}\pi^2$, to 80 decimal places.

\subsection{The odd moments $c_{6,2k+1}$}

Grouping the 6 internal lines of the 5-loop \emph{vacuum} diagram
\[V_6(a,b,c,d,e,f)=
\int_{a+b+c}^\infty \int_{d+e+f}^\infty
4uv\,D_4(a,b,c,u)V_2(u,v)D_4(v,d,e,f)\,{\rm d}v\,{\rm d}u\]
into two sets of 3 lines, we obtain
\begin{equation}
c_{6,1}=\int_{0}^\frac13D(y)\int_{0}^\frac13
\frac{D(z)\log(z/y)}{z^2-y^2}\,{\rm d}z\,{\rm d}y
\label{c61}
\end{equation}
after setting the masses to unity and transforming to $y=1/u$
and $z=1/v$. We also note that Broadhurst conjectured that
the value of $c_{6,5}$ is
$\frac{85}{72}c_{6,3}-\frac{1}{36}c_{6,1}+\frac{5}{48}$.
This was later checked to 500 decimal places, using
data in~\cite{ising-data}.

\subsection{The even moment $c_{6,0}$}

Finally, we obtain 3 complete
elliptic integrals in the integrand of
\begin{equation}
\overline{V}_6(a,b,c,d,e,f)=\pi\int_{-\infty}^\infty\int_{-\infty}^\infty
\overline{S}_3(a,b,u)\overline{S}_3(c,d,u+v)\overline{S}_3(e,f,v)
\,{\rm d}u\,{\rm d}v
\label{novee6}
\end{equation}
using the \emph{Aufbau}~(\ref{Aufbau}). This then delivers
\begin{equation}
c_{6,0}=\frac{\pi}{2}
\int_{-\pi/2}^{\pi/2}
\int_{-\pi/2}^{\pi/2}
\frac{{\bf K}(\sin\theta)\,{\bf K}(\sin\phi)\,
{\bf K}\left(\frac{\sin(\theta+\phi)}
{\sqrt{\cos^2\theta\cos^2\phi+\sin^2(\theta+\phi)}}\right)}
{\sqrt{\cos^2\theta\cos^2\phi+\sin^2(\theta+\phi)}}
\,{\rm d}\theta\,{\rm d}\phi
\label{cee60}
\end{equation}
by the same transformations as for~(\ref{cee50}).

\section{Computational notes}

This paper contains several proofs of identities that we first
conjectured on the basis of numerical investigation,
hugely facilitated by access to Sloane's wonderful sequence finder.
For the many one-dimensional integrals that we have noted,
we were greatly aided by the efficiency of the
{\tt agm} and {\tt intnum} procedures
of \emph{Pari-GP}, for evaluations of integrands and
integrals at precisions up to 1200 decimal places.
Results were then fed to the implementation of
the PSLQ algorithm in \emph{Pari-GP}'s {\tt lindep}
procedure, with which we performed many unsuccessful
searches for integer relations, as well as obtaining
the positive results reported in the paper.
By way of example, we remark that the integral
\[{\mathcal E}:=\int_{0}^\infty\frac{E(x)}{\sqrt{4+x^2}}\,
{\rm arcsinh}^2\left(\frac{x}{2}\right)\,{\rm d}x\]
was evaluated to high precision in order to
search for relations between $c_{5,1}$, $c_{5,3}$,
${\mathcal E}$ and products of powers of $\pi$,
$C$ and $1/C$, with coefficients that might
be Q-linear combinations of 1, $\sqrt{3}$,
$\sqrt{5}$ and $\sqrt{15}$. No such relation
was found.

\emph{Maple} was especially useful
for its {\tt HeunG}, {\tt MeijerG} and {\tt Gosper}
procedures and also for quick {\tt PSLQ} searches
with few terms, at relatively low precision.

But for two-dimensional numerical quadratures we
found neither \emph{Pari-GP} nor \emph{Maple}
to be remotely adequate for our demanding investigations.
For these, we came to rely on Bailey's multiprecision
code for two-dimensional integrals~\cite{hiprec,boxintegrals,quadrature},
which confirmed, to more than 100 decimal places,
the correctness of derivations of~(\ref{cee50}),~(\ref{cee60}) and
other identities. Here we offer a brief description of this scheme.

\subsection{Multi-dimensional quadrature}

Bailey's 1-D and 2-D schemes, as well as the one-dimensional {\tt intnum}
procedure in \emph{Pari-GP}, employ the tanh-sinh quadrature algorithm,
which was originally discovered by Takahasi and Mori~\cite{takahasi}. It is
rooted in the Euler-Maclaurin summation formula~\cite[p.\ 285]{atkinson},
which implies that for certain bell-shaped integrands $f(x)$ on $[0,1]$
where the function and all higher derivatives rapidly approach zero at the
endpoints, approximating the integral of $f(x)$ by a simple step-function
summation is remarkably accurate. This observation is combined with the
transformation $x = \tanh ((\pi/2)\sinh t)$, which converts most
``reasonable" integrand functions on $(-1,1)$ (including many functions
with singularities or infinite derivatives at one or both endpoints) into
bell-shaped functions with the desired property.

In particular, we can write, for an interval length $h > 0$,
\begin{equation}
\int_{-1}^1 f(x) \, {\rm d}x \; = \;
\int_{-\infty}^\infty f(g(t)) g^\prime(t) \, {\rm d}t
\; \approx \; h \sum_{j=-N}^N w_j f(x_j)\,,
\label{sum-form}
\end{equation}
where $g(t) = \tanh((\pi/2)\sinh t), \; x_j = g(h j), \; w_j = g'(h j)$,
and $N$ is chosen large enough that $|w_j f(x_j)| < \epsilon$
for $|j|>N$. Here $\epsilon = 10^{-p}$, where $p$ is the numeric precision
level in digits. Note that the resulting quadrature formula~(\ref{sum-form})
has the form similar to Gaussian quadrature, namely a simple summation
with abscissas $x_j$ and weights $w_j$, both sets of which can be
pre-computed since they are independent of the integrand function.
For many integrand functions, once $h$ is sufficiently small, reducing
$h$ by half yields twice as many correct digits in the result (although
all computations must be performed to at least the level of precision
desired for the final result, and perhaps double this level if the function
is not well-behaved at endpoints). Additional details of efficient
implementations are given in~\cite{hiprec,quadrature}.

One of the numerous applications of one-dimensional tanh-sinh quadrature
in this study was the verification of our final conjecture given
in~(\ref{conj11}). This was
done using Bailey's implementation of the one-dimensional tanh-sinh
algorithm, together with the ARPREC extreme-precision software package
\cite{ARPREC}. Evaluation of the Bessel function $I_0(t)$ was performed
using a hybrid scheme where formula 9.6.12 of~\cite{AandS} was used for
modest-sized values, and formula 9.7.1 for large values. Evaluation of
$K_0(t)$ was performed using 9.6.13 of~\cite{AandS} for modest-sized
values, and 9.7.2 for large values. Note however that formula 9.6.13
for $K_0(t)$ must be implemented using a working precision that is
roughly twice the level desired for the final result, due to the
sensitive subtraction operation in this formula. Also note that when $m
= 0$ in~(\ref{conj11}), this combination of formulas is not
satisfactory, because for large $t$ the function $I_0(t)$ is
exceedingly large, and $K_0(t)$ is exceedingly small, and even though
the product is of modest size, overflows and underflows are possible in
intermediate function evaluations, even when using high-precision
software that has an enormously extended dynamic range. For such cases
($m = 0$ and $t$ large), we employed formula 9.7.5 of~\cite{AandS},
which gives an asymptotic series for the product $I_0(t) K_0(t)$.
We had differently addressed this issue
in the special case of~(\ref{u61alt}).

Armed with an efficient implementation of these schemes, we were able to
verify~(\ref{conj11}) for all $(n,k)$ pairs, where $1 \leq k \leq
\lfloor n/2 \rfloor$ and $4 \leq n \leq 12$ (there are 43 such pairs),
in each case to over 340-digit accuracy. In addition, we verified
(\ref{conj11}) for a variety of larger $(n,k)$ pairs, including $(15,7),
\; (20,10), \; (25, 11), \; (30,12), \; (37,13), \;$ and $(41,14),$ again
to over 340-digit accuracy in each case.

The tanh-sinh quadrature algorithm can also be performed in two
or more dimensions as an iterated version of the one-dimensional
scheme. Such computations are many times more expensive than
in one dimension. For example, if roughly 1,000 function evaluations
are required in one dimension to achieve a desired precision level,
then at least 1,000,000 function evaluations are typically required in
two dimensions, and 1,000,000,000 in three dimensions.
Additionally, the behaviour of multi-dimensional tanh-sinh quadrature
on integrand functions with singularities or infinite derivatives on the
boundaries of the region is not as predictable or well-understood as in
one dimension.

\begin{figure}
\centerline{\hss
\epsfig{file=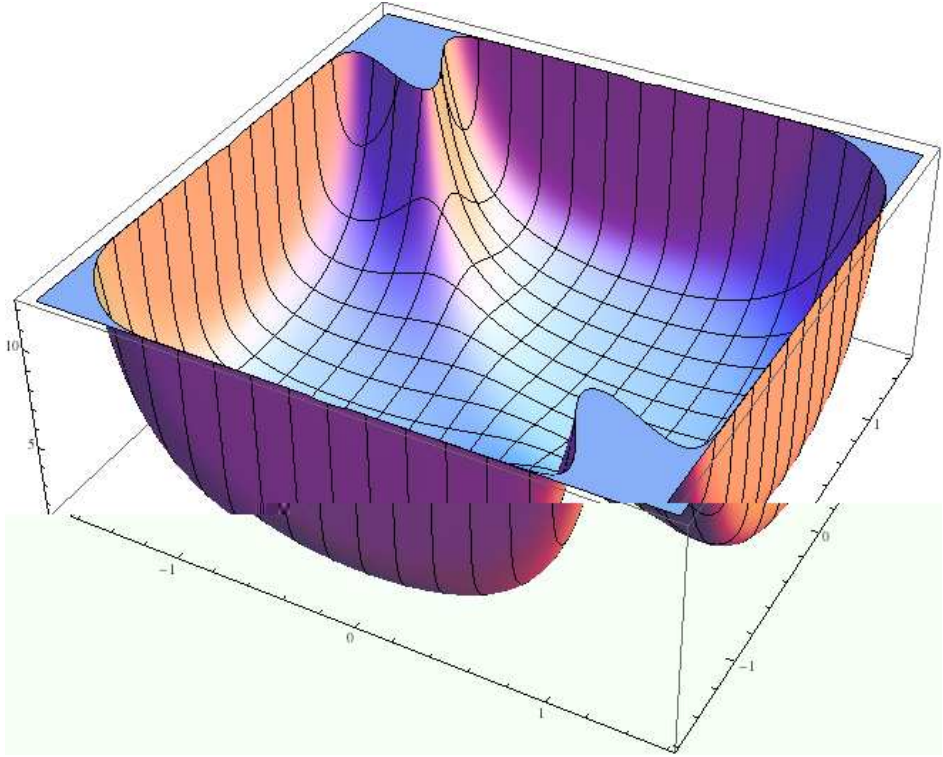,height=10cm}\hss}
\caption{Plot of $c_{5,0}$ integrand function in~(\ref{cee50-1})}
\end{figure}

Nonetheless, we were able to use 2-D tanh-sinh quadrature to
successfully evaluate a number of the double integrals mentioned in
this paper, after making some minor transformations. As one example,
consider the integral mentioned above for $c_{5,0}$, namely
\begin{equation}
c_{5,0}=\frac{\pi}{2}
\int_{-\pi/2}^{\pi/2}
\int_{-\pi/2}^{\pi/2}
\frac{{\bf K}(\sin\theta)\,{\bf K}(\sin\phi)}
{\sqrt{\cos^2\theta\cos^2\phi+4\sin^2(\theta+\phi)}}
\,{\rm d}\theta\,{\rm d}\phi\,.
\label{cee50-1}
\end{equation}

Note that this function (see Figure 1) has singularities on all four sides
of the region of integration, with particularly troublesome singularities
at $(\pi/2,-\pi/2)$ and $(-\pi/2, \pi/2)$. However, after making the
substitutions $s \leftarrow \pi/2 - s, \; t \leftarrow \pi/2 - t$ and
$r \leftarrow s/t$, and taking advantage of the symmetry evident from
Figure 1, we obtain
\begin{eqnarray}
c_{5,0} &=& 2 \pi \int_0^{\pi/2} \int_0^1
 \frac{t\,{\bf K}(\cos (r t))\,{\bf K}(\cos t)\, {\rm d} r\, {\rm d} t}
 {\sqrt{\sin^2 (r t) \sin^2 t + 4 \sin^2 (t(1 + r))}} \nonumber \\
&& {}+ 2 \pi \int_0^{\pi/2} \int_0^1
 \frac{t\,{\bf K}(\cos (r t))\,{\bf K}(\cos t)\, {\rm d} r\, {\rm d} t}
 {\sqrt{\sin^2 (r t) \sin^2 t + 4 \sin^2 (t(1 - r))}}\,,
\label{cee50-2}
\end{eqnarray}
which is significantly better behaved (although these integrands
still have singularities on two of the four sides of the region). As a
result, we were able to compute $c_{5,0}$ with this formula to
120-digit accuracy, using 240-digit working precision. This run
required a parallel computation (using the MPI parallel
programming library) of 43.2 minutes on 512 CPUs (1024 cores)
of the ``Franklin" system at the National Energy Research Scientific
Computing Center at the Lawrence Berkeley National Laboratory.
The final result matched the value that we had previously
calculated using~(\ref{moments}) (see~\cite{ising-data}) to
120-digit accuracy.

This same strategy was successful for several other 2-D integrals.
For example, we computed $c_{6,0}$ to 116-digit accuracy, which
again matched the value we had previously computed, in 64.8
minutes on 1024 cores of the Franklin system. In the case of
$c_{6,1}$, the transformation described above for $c_{5,0}$
converted the integrand function of~(\ref{c61}) into a completely
well-behaved function, without any singularities. As a result, we
were able to compute $c_{6,1}$ to 120-digit accuracy using only
an Apple Intel-based workstation with four computational cores,
in 28 minutes. As before, the result matched the earlier calculation.

As already noted, complex numbers are avoided in
integral~(\ref{s61}) by writing it as
\begin{equation}
s_{6,1}=\int_0^\frac13D(y)\int_0^\frac12
\frac{4z \, f(A_-/A_+)}{A_+ \sqrt{1-4z^2}}
\,{\rm d}z\,{\rm d}y
\label{s61-rev}
\end{equation}
where
\[f(x) := \left\{ \begin{array}{lll}
{\rm arctanh}\left(\sqrt{x}\right)/\sqrt{x} &\mbox{ for }& x > 0\\
1 &\mbox{ for }& x = 0\\
\arctan\left(\sqrt{-x}\right)/\sqrt{-x} &\mbox{ for }& x < 0
\end{array} \right.\]
yields positive real numbers within the rectangle of integration.
We were able to confirm that the double integral~(\ref{s61-rev})
yields the first 120 of the 1200 decimal places obtained,
far more easily, from the single integral~(\ref{s61conj}).

\section{Conclusions}

Despite some notable progress in discovering and proving new
results, we are left with 8 outstanding conjectures.\footnote{The
conjecture in (\ref{DJB3a}) was later proven by contour integration --- see 
\cite{djb-2008a}.}   Of these,
5 have their first instances in Equations~(\ref{Lap})
to~(\ref{s55}) and Equation~(\ref{DJB3a}) of Section~5.1,
with 3 outliers, in Equations~(\ref{c55}),~(\ref{DJBrelE3})
and~(\ref{conj11}).

Conjecture~(\ref{Lap}) lies deep in 4-dimensional quantum field theory,
but it is reasonable to suppose that it might be derivable from
the two-dimensional conjectures~(\ref{s51}),~(\ref{s53})
and~(\ref{s55}), together with their proven sign-changed variants
in~(\ref{t51}),~(\ref{t53}) and~(\ref{t55}).

The conjectural integer relations~(\ref{c55}) and~(\ref{s55rec})
may be provable by adding rational data such
as~(\ref{trivial}) to a set of recursions richer than those
considered in~\cite{bsalvy,ouvry}.

The real challenge is set by the remarkable sum rule~(\ref{sum}),
with a dispersive presentation~(\ref{sumD}),
a non-dispersive presentation~(\ref{sumE}) and
a kindergarten analogue~(\ref{DJBrelE3}).

\paragraph{Acknowledgements.} We are most grateful to Andrei
Davydychev, for elucidation of dilogarithms obtained in quantum
field theory, to Stefano Laporta, for the astute observations in
4-dimensional quantum field theory from which our work arose, to
Bas Tausk, for advice on Mellin--Barnes transforms, and to John
Zucker, for several pertinent observations regarding integrals of
elliptic integrals.

\newpage

\raggedright

\end{document}